\documentclass[acmtog,authorversion]{acmart}

\newcommand{\final}{1}

\usepackage{acmart-taps}

\usepackage{color}
\usepackage{ifthen}
\usepackage{float}
\usepackage{alltt}
\usepackage{newlfont} %
\usepackage{floatflt}
\usepackage{wrapfig}
\usepackage{fixltx2e}
\usepackage{subfig} %
\usepackage{multirow}
\usepackage{booktabs}
\usepackage{cleveref}
\usepackage{algorithmic}
\usepackage{mathtools, cuted}
\usepackage{CJKutf8} %

\usepackage[ruled]{algorithm2e} %

\usepackage{tikz}

\setcitestyle{square}

\SetAlFnt{\small}
\SetAlCapFnt{\small}
\SetAlCapNameFnt{\small}
\SetAlCapHSkip{0pt}
\IncMargin{-\parindent}
\newcommand{\Caption}[2]{\caption[#1]{{\em #1} #2}}

\newcommand{\emacsquote}[1]{{``#1''}}

\definecolor{SithColor}{rgb}{0.7,0,0} %
\newcommand{\liyi}[1]{{\color{SithColor} Li-Yi: #1 $\qed$}}
\definecolor{ConsularColor}{rgb}{0,0.4,0} %
\definecolor{GuardianColor}{rgb}{0,0,0.8} %
\newcommand{\peihan}[1]{{\color{GuardianColor} Peihan: #1 $\qed$}}
\newcommand{\yoda}[1]{{\color{ConsularColor} Yoda: #1 $\qed$}}
\newcommand{\koji}[1]{{\color{ConsularColor} Koji: #1 $\qed$}}

\definecolor{MatthiasColor}{rgb}{0,0.7,0} 
\newcommand{\matthias}[1]{{\color{MatthiasColor} Matthias: #1 $\qed$}}

\newcommand{\warning}[1]{{\it\color{red} #1}}
\newcommand{\note}[1]{{\it\color{blue} #1}}
\newcommand{\nothing}[1]{}

\definecolor{AudioColor}{rgb}{0.56,0.34,0.62}

\definecolor{VideoColor}{rgb}{0.44,0.66,0.38}

\definecolor{NewColor}{rgb}{0.9,0.4,0}
\newcommand{\new}[1]{{\color{NewColor} #1}}
\definecolor{DeleteColor}{rgb}{0.1,0.6,1.0}
\newcommand{\delete}[1]{{\color{DeleteColor} #1}}
\definecolor{MoveColor}{rgb}{0.5,0.1,0.5}

\newcommand{\replace}[2]{\delete{#1}\new{#2}}

\definecolor{FinalNewColor}{rgb}{0.9,0.4,0}
\newcommand{\finalnew}[1]{{\color{NewColor} #1}}

\definecolor{figred}{rgb}{1,0,0}
\definecolor{figgreen}{rgb}{0,0.6,0}
\definecolor{figblue}{rgb}{0,0,1}
\definecolor{figpink}{rgb}{1,0.63,0.63}

\ifthenelse{\isundefined{\hili}}
{
\renewcommand{\new}[1]{{#1}}
\renewcommand{\delete}[1]{}

\renewcommand{\finalnew}[1]{{#1}}
}
{
\ifthenelse{\equal{\hili}{0}}
{
\renewcommand{\new}[1]{{#1}}
\renewcommand{\delete}[1]{}

\renewcommand{\finalnew}[1]{{#1}}
}
{}
}

\ifthenelse{\equal{\final}{1}}
{
\renewcommand{\liyi}[1]{}
\renewcommand{\peihan}[1]{}
\renewcommand{\yoda}[1]{}
\renewcommand{\koji}[1]{}
\renewcommand{\matthias}[1]{}
\renewcommand{\warning}[1]{}
\renewcommand{\note}[1]{}
}
{}

\newcommand{\pseudocode}{Algorithm}
\floatstyle{plain}
\newfloat{algorithm}{tbhp}{lop}
\floatname{algorithm}{\pseudocode}

\newcommand{\filename}[1]{\url{#1}}
\newcommand{\foldername}[1]{\url{#1}}

\definecolor{darkgreen}{RGB}{0,100,0} 
\definecolor{skyblue}{RGB}{135,206,235} %

\newtheorem{textonprop}{Property}

\newcommand{\norm}[1]{\lVert #1 \rVert}

\newcommand{\imageappearancefeaturedim}{d_{a}}

\newcommand{\gaussianencoder}{\mathcal{E}}
\newcommand{\fcencoder}{\gaussianencoder_{FC}} %

\newcommand{\segmentationnetwork}{\gaussianencoder_s}
\newcommand{\gaussianestimationlayer}{\gaussianencoder_g}

\newcommand{\gaussiangenerator}{\mathcal{G}}
\newcommand{\discriminator}{\mathcal{D}}

\newcommand{\image}{I}
\newcommand{\imagetransformed}{\hat{\image}}

\newcommand{\imagerecon}{\image_{r}}

\newcommand{\imagereshufflerecon}{\tilde{\image}_{r}}
\newcommand{\imageheight}{H}
\newcommand{\imagewidth}{W}
\newcommand{\imagefeature}{\mathbf{F}}
\newcommand{\imageappearancefeature}{\imagefeature_a}

\newcommand{\imagedirfeature}{\mathbf{V}}

\newcommand{\segmap}{\mathbf{S}}

\newcommand{\segment}{\segmap_\gaussianindex}
\newcommand{\normsegmap}{\mathfrak{S}_\gaussianindex}
\newcommand{\realnumberset}{\mathbb{R}}

\newcommand{\segmentareasymbol}{A}
\newcommand{\segmentarea}{\segmentareasymbol_{\gaussianindex}}

\newcommand{\interpcoe}{\eta}

\newcommand{\latent}{\mathbf{z}}

\newcommand{\gaussiansymbol}{\textbf{g}}
\newcommand{\gaussian}{\gaussiansymbol}

\newcommand{\gaussianreshuffle}{\tilde{\gaussian}}

\newcommand{\gaussiancomposition}{\setof{\gaussian}{\gaussianindex}_{\gaussianindex=1}^{\numgaussians}}

\newcommand{\gaussiancompositionreshuffle}{\setof{\gaussianreshuffle}{\gaussianindex}_{\gaussianindex=1}^{\numgaussians}}

\newcommand{\setof}[2]{\{#1_{#2}\}}
\newcommand{\gaussianindex}{i}
\newcommand{\gaussianindexj}{j}
\newcommand{\numgaussians}{n}

\newcommand{\gaussianweightsymbol}{\delta}
\newcommand{\gaussianweight}{\gaussianweightsymbol_\gaussianindex}

\newcommand{\gaussianweightreshuffle}{\tilde{\gaussianweightsymbol}_\gaussianindexj}

\newcommand{\gaussiandirsymbol}{\mathbf{\nu}}
\newcommand{\gaussiandir}{\gaussiandirsymbol_\gaussianindex}

\newcommand{\gaussianmeansymbol}{\boldsymbol{\mu}}
\newcommand{\gaussianfeaturesymbol}{\mathbf{f}}

\newcommand{\gaussianmean}{\gaussianmeansymbol_\gaussianindex}

\newcommand{\gaussianfeature}{\gaussianfeaturesymbol_\gaussianindex}

\newcommand{\gaussianfeaturesymbolfunc}[1]{\gaussianfeaturesymbol_{#1}}

\newcommand{\gaussianinterp}{\gaussiansymbol_\gaussianindex^\dagger}

\newcommand{\gaussianfeaturevar}{\Delta_\gaussianfeaturesymbol}

\newcommand{\gaussianfeaturedim}{n_f}

\newcommand{\gaussiancovariancesymbol}{\mathbf{U}}
\newcommand{\gaussiancovariance}{\gaussiancovariancesymbol_\gaussianindex}

\newcommand{\gaussiancovariancevar}{\Delta_\gaussiancovariancesymbol}

\newcommand{\xcoord}{x}
\newcommand{\ycoord}{y}
\newcommand{\coordvec}{\mathbf{p}}
\newcommand{\reconcoordvec}{\mathbf{p}^\prime}

\newcommand{\bernoullidistribution}{B}
\newcommand{\bernoulliprob}{p}
\newcommand{\gaussianexistencesymbol}{\bernoulliprob}
\newcommand{\gaussianexistence}{\gaussianexistencesymbol_{\gaussianindex}}

\newcommand{\gaussianexistencereshuffle}{\tilde{\gaussianexistencesymbol}_{\gaussianindexj}}

\newcommand{\gaussiansplatting}{h}
\newcommand{\gaussiansplattedfeature}{\mathbf{F}_{gs}}

\newcommand{\splatalpha}{\alpha}

\newcommand{\generatorinputfeature}{\mathbf{F}_{input}}

\newcommand{\reshufflefactor}{\tau_{\gaussianindex\rightarrow\gaussianindexj}}
\newcommand{\reshufflepower}{\gamma}

\newcommand{\geometrictransform}{T}

\newcommand{\gaussiantransformed}{\hat{\gaussian}}

\newcommand{\reshufflesymbol}{\pi}

\newcommand{\loss}{L}
\newcommand{\lossrecon}{\loss_{R}}
\newcommand{\lossreconprime}{\loss^{\prime}_{R}}

\newcommand{\lossreconnorm}{\loss_{1}}
\newcommand{\losslpips}{\loss_{LPIPS}}
\newcommand{\lossgan}{\loss_{GAN}}
\newcommand{\losspatchgan}{\loss_{PGAN}}

\newcommand{\weightreconnorm}{w_{1}}
\newcommand{\weightlpips}{w_{LPIPS}}

\newcommand{\lossentropy}{\loss_{E}}

\newcommand{\losscompact}{\loss_{C}}
\newcommand{\lossmatch}{\loss_{Csis}}
\newcommand{\losstexture}{\loss_{T}}

\newcommand{\weightentropy}{w_{E}}

\newcommand{\weightcompact}{w_{C}}
\newcommand{\weightmatch}{w_{Csis}}
\newcommand{\weighttexture}{w_{T}}
\newcommand{\weightgan}{w_{GAN}}
\newcommand{\weightpatchgan}{w_{PGAN}}

\newcommand{\matching}{\sigma}

\newcommand{\autoencodingbox}{\dashedbox{red}{autoencoding}}
\newcommand{\reshufflingbox}{\dashedbox{darkgreen}{reshuffling}}
\newcommand{\disentanglingbox}{\dashedbox{skyblue}{disentangling}}

\newcommand{\dashedbox}[2]{%
    \tikz[baseline=(boxed word.base)]{
        \node[inner sep=0.8mm, dashed, draw={#1}] (boxed word) {#2};
    }
}

\copyrightyear{2024}
\acmYear{2024}
\setcopyright{acmlicensed}\acmConference[SA Conference Papers '24]{SIGGRAPH Asia 2024 Conference Papers}{December 3--6, 2024}{Tokyo, Japan}
\acmBooktitle{SIGGRAPH Asia 2024 Conference Papers (SA Conference Papers '24), December 3--6, 2024, Tokyo, Japan}
\acmDOI{10.1145/3680528.3687561}
\acmISBN{979-8-4007-1131-2/24/12}

\setcopyright{acmlicensed}

\begin{document}

\title{Compositional Neural Textures} %

\author{Peihan Tu}
\affiliation{
\institution{University of Maryland}
\city{College Park}
\state{Maryland}
\country{United States}
}%
\email{peihan.tu@gmail.com}

\author{Li-Yi Wei}
\affiliation{
\institution{Adobe Research}
\city{San Jose}
\state{California}
\country{United States}
}%
\email{review@liyiwei.org}

\author{Matthias Zwicker}
\affiliation{
\institution{University of Maryland}
\city{College Park}
\state{Maryland}
\country{United States}
}%
\email{zwicker@umd.edu}

\begin{abstract}

Texture plays a vital role in enhancing visual richness in both real photographs and computer-generated imagery. 
However, the process of editing textures often involves laborious and repetitive manual adjustments of textons, which are the recurring local patterns that characterize textures. 
This work introduces a fully unsupervised approach for representing textures using a compositional neural model that captures individual textons.
We represent each texton as a 2D Gaussian function whose spatial support approximates its shape, and an associated feature that encodes its detailed appearance.
By modeling a texture as a discrete composition of Gaussian textons, the representation offers both expressiveness and ease of editing.
Textures can be edited by modifying the compositional Gaussians within the latent space, and new textures can be efficiently synthesized by feeding the modified Gaussians through a generator network in a feed-forward manner.
This approach enables a wide range of applications, including transferring appearance from an image texture to another image, diversifying textures, texture interpolation, revealing/modifying texture variations, edit propagation, texture animation, and direct texton manipulation.
The proposed approach contributes to advancing texture analysis, modeling, and editing techniques, and opens up new possibilities for creating visually appealing images with controllable textures.

\end{abstract}


\begin{teaserfigure}
  \centering
  \setlength{\tabcolsep}{0.1cm}
  \begin{tabular}{ccc}
    \multicolumn{3}{c}{
    \subfloat[Overview]{%
    \label{fig:teaser:overview}%
    \includegraphics[width=0.95\linewidth]{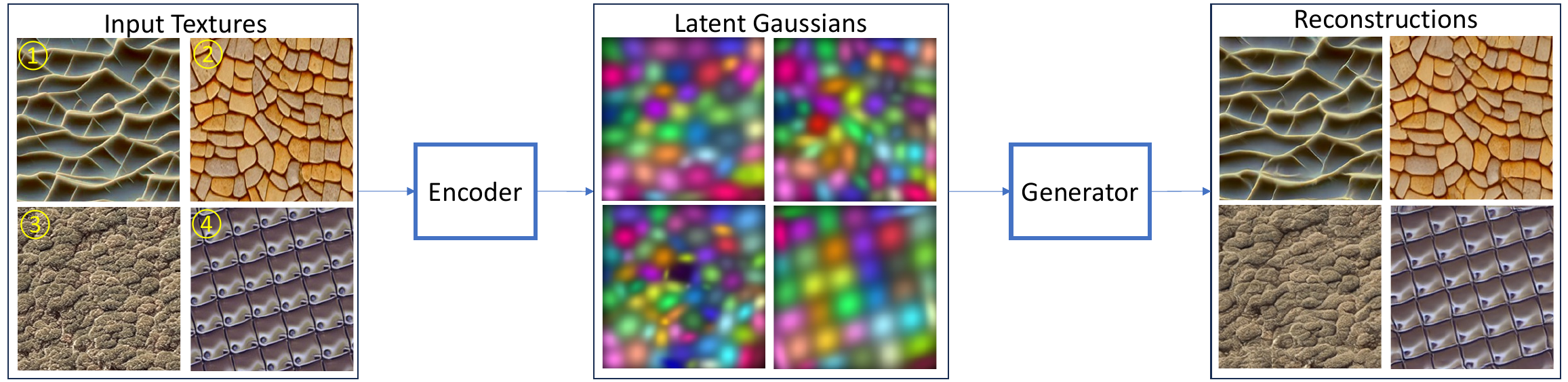}
    }%
    }
    \\[-0.3cm]
    \subfloat[Texture transfer(struct./app.$\textcircled{1}/\textcircled{2}$and$\textcircled{2}/\textcircled{1}$)]{%
    \label{fig:teaser:texture_transfer}%
    \includegraphics[width=0.15\linewidth]{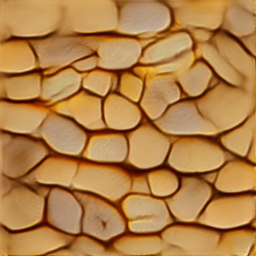}
    \includegraphics[width=0.15\linewidth]{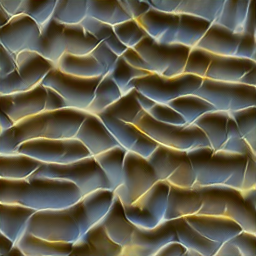}
    }%
    &
    \subfloat[Image stylization $\textcircled{1}$]{%
    \label{fig:teaser:image_stylization}%
    \includegraphics[width=0.15\linewidth]{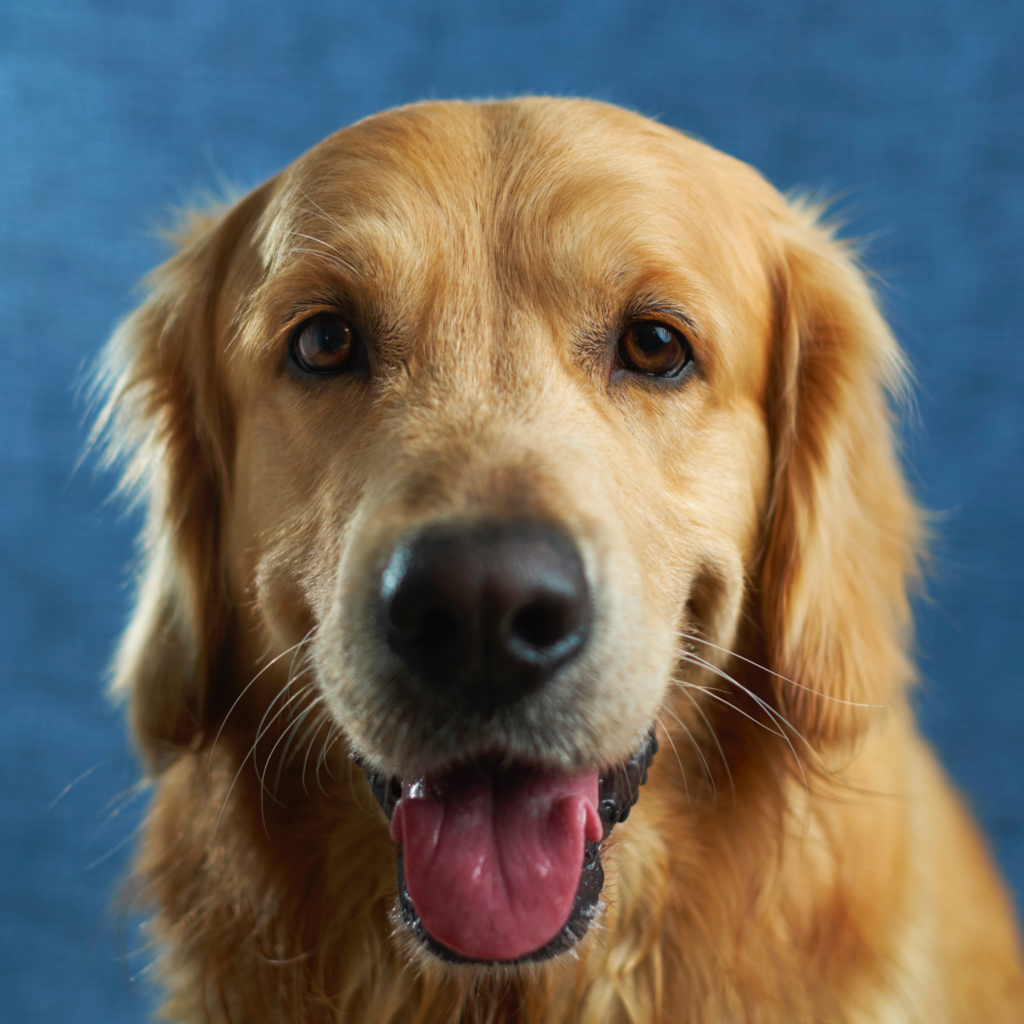}
    \includegraphics[width=0.15\linewidth]{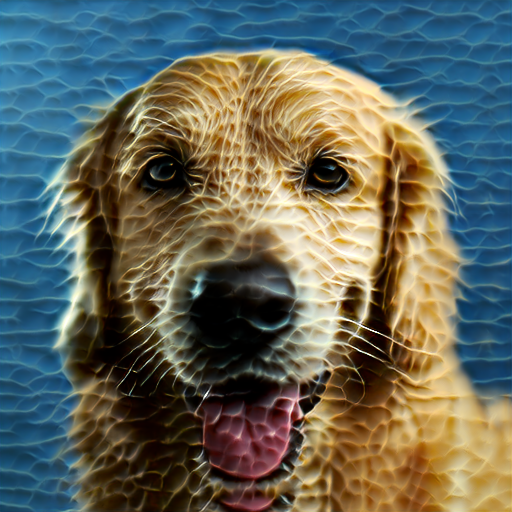}
    }%
    &
    \subfloat[Interpolation \& morphing $\textcircled{1}\textcircled{2}$]{%
    \label{fig:teaser:interp_morphing}%
    \includegraphics[width=0.15\linewidth]{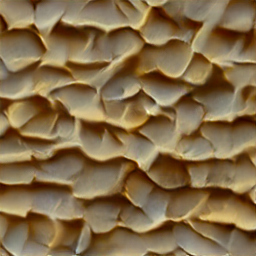}
    \includegraphics[width=0.15\linewidth]{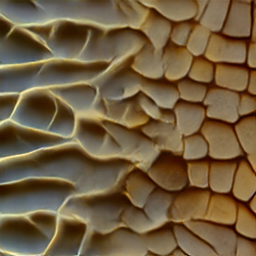}
    }%
    \\[-0.2cm]
    \subfloat[Revealing/modifying variations $\textcircled{3}$]{%
    \label{fig:teaser:variations}%
    \includegraphics[width=0.15\linewidth]{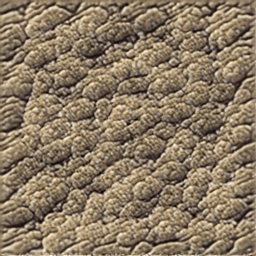}
    \includegraphics[width=0.15\linewidth]{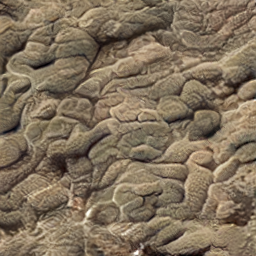}
    }%
    &
    \subfloat[Edit propagation $\textcircled{4}$]{%
    \label{fig:teaser:propagation}%
    \includegraphics[width=0.15\linewidth]{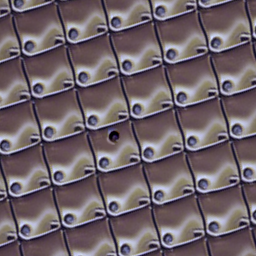}
    \includegraphics[width=0.15\linewidth]{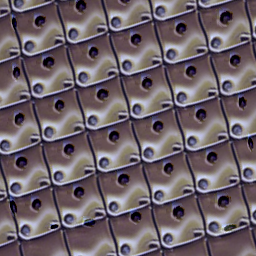}
    }%
    &
    \subfloat[Editing Gaussian textons $\textcircled{4}$]{%
    \label{fig:teaser:gaussians}%
    \includegraphics[width=0.15\linewidth]{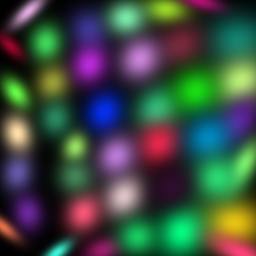}
    \includegraphics[width=0.15\linewidth]{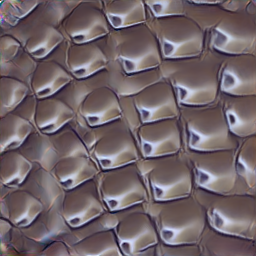}
    }%
  \end{tabular}
  \caption[]{\textit{Compositional neural representation for texture editing.} (a) We present a neural representation that decomposes textures into geometric elements with appearance features in a fully unsupervised manner.
Geometric elements are represented by Gaussian functions corresponding to individual texture elements (textons), and each Gaussian has an associated appearance vector to encode texton appearance.
 Our network architecture and objectives are designed to encourage extraction of recurring textons and separation of geometry/structure and appearance, while preserving faithful reconstruction of the original texture.
 This facilitates a variety of editing operations and synthesis applications, such as transferring appearance from one texture towards the structure of another texture (b) or image (c), texture interpolation and morphing (d), manipulating the variations within a texture (e), edit propagation (f), and direct manipulation (e.g. moving and transforming) of the textons (g). Copyright: (c) (left) by Victor G from Unsplash; unless copyrighted, input images are generated from Adobe Firefly throughout the paper.
 }
 \label{fig:teaser}
\end{teaserfigure}

\maketitle

\section{Introduction}
\label{sec:introduction}

Texture plays a vital role in enhancing visual richness in both real photographs and computer-generated imagery by providing intricate details, patterns, and surface characteristics.
In practice, users often need to create or adjust textures in an image, but the process can be challenging and time-consuming, often requiring laborious manual adjustments due to the repetitive details.
To avoid ambiguity, we follow the definition of \emacsquote{textures} as repetitive patterns \cite{Wei:2009:SAE} instead of general images mapped to 3D objects \replace{\cite{Siddiqui:2022:GT3,Chen:2022:AUV,Cao:2023:TF,Kim:2023:PI}}{\cite{Blinn:1976:TR,Heckbert:1986:STM,Cao:2023:TF,Chen:2022:AUV}}.

Previous research in texture manipulation has explored various applications, ranging from global control \cite{Hertzmann:2001:IA,Rosenberger:2009:LSS,Kaspar:2015:STT},
regional variations \cite{Matusik:2005:TDU,Fruhstuck:2019:TG,Yu:2019:TM,Bellini:2016:TVW,Dekel:2015:RMN}, to individual textons \cite{Brooks:2002:SSB}.
These approaches rely on domain-specific priors for different applications or categories of textures.
Ideally, we would like to have a general-purpose approach that can be applied to a wide range of textures and editing tasks across different scales.

In this paper, we introduce a texture representation that decomposes textures into a composition of recurring neural elements (Figure~\ref{fig:teaser:overview}), disentangling the geometry and appearance of each element. We obtain this representation by developing an autoencoder architecture that is trained in a fully unsupervised manner.
Our representation is human-intelligible, applicable to a wide range of textures that exhibit sufficient self-repetitions, and enhancing various editing techniques (Figure~\ref{fig:teaser}).
Textures can be edited by manipulating the compositional neural representations in a deep latent space, and \delete{edited textures can be}efficiently generated by feeding the manipulated representations through a generator (decoder) network in a feed-forward manner.
The supported editing operations include, but are not limited to,
transferring appearance from a texture to another image (Figures~\ref{fig:teaser:texture_transfer} and \ref{fig:teaser:image_stylization}), diversifying textures (Figure~\ref{fig:diverse_texture_synthesis_main}), texture interpolation and morphing (Figure~\ref{fig:teaser:interp_morphing}), revealing/modifying texture variations (Figure~\ref{fig:teaser:variations}), edit propagation (Figure~\ref{fig:teaser:propagation}), texture animation (supplementary video), and direct manipulation of textons (Figure~\ref{fig:teaser:gaussians}).

Our main idea of representing textures using a compositional neural model is rooted in the understanding that textures are characterized by their recurring patterns, which can vary from regular to irregular formations. Within each texture, basic recurring patterns are known as textons, acting as building blocks that define the texture's unique appearance and structure.
However, extracting textons from diverse textures poses considerable challenges due to their ambiguous nature and the difficulty in defining them rigorously.
One possible texton extraction approach is to use supervised learning, for example, by training image segmentation networks.
While supervised approaches have achieved state-of-the-art performance for extracting well-defined concepts from images \cite{Krizhevsky:2012:ICD,Minaee:2021:ISD} using massive, annotated datasets,
these are less suitable for extracting textons due to their numerous and ambiguous nature that can be challenging for consistent annotations with reasonable quality, cost, and time.
Therefore, we develop a fully unsupervised learning approach to extract neural textons.

First, we propose a set of desired properties of (neural) textons detailed in Section~\ref{sec:method:properties}, \delete{which we leverage to effectively train the autoencoder (Figure~\ref{fig:teaser:overview}),
}%
including
1) discrete objectness, 2) spatial compactness, 3) reshuffling invariance and 4) transformation equivariance.
These properties enable the textons to represent diverse sets of textures while maintaining human interpretability for intuitive texture editing.
\new{As per these properties}, second, 
we represent each neural texton as a 2D Gaussian with an associated appearance feature.
A Gaussian approximates a texton's shape via its spatial support, and the appearance feature encodes the texton's exact appearance.
This separation facilitates the enforcement of geometric transformation equivariance.
Further, the separate encoding of rough spatial and detailed appearance information in Gaussian covariances and associated features offers comprehensibility and intuitive user control over textons \cite{Hertz:2022:EIS,Epstein:2022:BGAN}. 
Third, 
we develop unsupervised training methods to instill the desired properties in neural textons by using proper loss functions.
For example, the reshuffling invariance property dictates that randomly reshuffling the appearances among textons should not modify the texture's overall appearance.
We thus design two training losses to assess the texture similarity between the input texture and a reconstructed texture with reshuffled texton appearance features.
Finally,
we observe that the training losses may present contradictory behaviors, where minimizing one could lead to the increase of another, and address this issue by carefully adjusting the training schedules to achieve a balanced trade-off between these losses.

Our main contributions are: 1) proposing the first compositional neural representation for textures;
2) demonstrating its versatility for supporting various texture editing applications without additional model training;
3) illustrating our approach's unique capabilities in modeling various textures as textons through comparisons and ablation studies;
and 4) planning to release our code and models to facilitate future research in texture editing and representation.

\section{Related Work}
\label{sec:prior}

\paragraph{Neural Texture Representations}

Recent advancements in deep texture representations have yielded impressive results in various applications, including texture classification
\cite{Zhai:2020:DSR,Xue:2018:DTM,Zhang:2017:TEN}, segmentation \cite{Minaee:2021:ISD,Li:2022:STN}, synthesis \cite{Gatys:2015:Texture,Zhou:2018:NST}, interpolation \cite{Yu:2019:TM,Bergmann:2017:LTM,Henzler:2020:LN3} and rectification \cite{Hao:2023:DHT}.
In addition, recent large generative models \cite{Rombach:2022:HRI,Karras:2020:AII} have significantly enhanced image representation and generation capabilities in general.
However, these representations, while engineered to encode a broad spectrum of appearance information, exhibit very limited or no disentanglement when applied to perception and uncontrolled synthesis tasks.
This entanglement considerably limits their practical utility in texture editing tasks.
Texture interpolation, a global texture editing task, necessitates a representation that effectively disentangles local and global texture appearances \new{\cite{Bergmann:2017:LTM,Yu:2019:TM,Henzler:2020:LN3}}.
\note{
Bergmann et al. \cite{Bergmann:2017:LTM} introduce a periodic spatial GAN (PSGAN). PSGAN maps noise to a texture through noise dimensions that are disentangled into global, local, and spatially periodic components. 
Yu et al. \shortcite{Yu:2019:TM} develop an autoencoder to encode an image into a disentangled latent space.
This space is divided into components that separately represent local and global aspects of texture appearance.
Specifically, the local components of these representations encode the stochastic variations within a texture, while the global components capture the overall appearance.
This separation makes these representations particularly effective for texture interpolation tasks.
It facilitates the interpolation of the overall texture appearance, without compromising the inherent stochasticity within textures.
Henzler et al. \shortcite{Henzler:2020:LN3} model stochastic textures using noise fields and an MLP decoder.
Similarity to \cite{Bergmann:2017:LTM,Yu:2019:TM}, this approach effectively disentangles local stochastic and global appearance components.
It achieves this by encoding stochasticity with noise fields and the overall appearance with a global texture code.
}%
\note{This disentanglement facilitates applications in texture interpolation.}%
However, the level of disentanglement in the representations currently used for interpolation is still inadequate, limiting their extension to other applications across a diverse range of textures.
To address this shortfall and broaden the scope in texture editing, a more effectively disentangled representation is essential.
We propose a compositional neural texture representation by modeling textures as neural textons in a deep latent space with disentangled structure and appearance.
The proposed compositional representation is applicable to a number of texture editing applications and a wide variety of textures.

\paragraph{Texture Editing}
\label{sec:prior:texture:edit}

Previous research in texture manipulation has explored various applications, such as controlled synthesis \cite{Hertzmann:2001:IA,Rosenberger:2009:LSS,Kaspar:2015:STT},
interpolation \cite{Matusik:2005:TDU,Fruhstuck:2019:TG,Yu:2019:TM,Bellini:2016:TVW}, revealing/modifying variations \cite{Dekel:2015:RMN}, edit propagation \cite{Brooks:2002:SSB}, texture weathering \cite{Bellini:2016:TVW} \new{and splicing \cite{Liu:2009:TS}}.
These approaches rely on domain-specific priors for specific applications or categories of textures.
They leverage image smoothness \cite{Kwatra:2003:GT} or patch similarity priors \cite{Lefebvre:2005:PCT,Zhou:2023:NTS}, procedural noise \cite{Henzler:2020:LN3} defined in image pixel space \cite{Lefebvre:2005:PCT,Brooks:2002:SSB}, hand-crafted feature space \cite{Lefebvre:2006:ATS,Liu:2004:NRT,Matusik:2005:TDU}, or learned feature spaces \cite{Fruhstuck:2019:TG,Yu:2019:TM,Bergmann:2017:LTM,Henzler:2020:LN3}, for local \cite{Wei:2009:SAE}, stochastic \cite{Henzler:2020:LN3}, regular/near-regular \cite{Liu:2004:NRT} or non-stationary textures \cite{Zhou:2018:NST}.

These texture editing methods can be classified into three categories: non-parametric \cite{Kwatra:2003:GT,Dekel:2015:RMN,Wei:2009:SAE,Lefebvre:2006:ATS}, parametric \cite{Matusik:2005:TDU,Liu:2004:NRT} and more recent deep learning-based approaches \cite{Yu:2019:TM,Henzler:2020:LN3}.
Non-parametric methods, known for their flexibility, can adapt to a diverse range of texture types.
However, their editing results are less predictable, and these methods are typically confined to specific applications \cite{Wei:2009:SAE,Bellini:2016:TVW,Brooks:2002:SSB,Dekel:2015:RMN,Lu:2009:DTD}.
In contrast, parametric methods are anchored in well-defined mathematical models, which makes them more interpretable, predictable, and controllable.
However, these methods encounter difficulties when dealing with textures that are rich in complexity, encompassing intricate details and sophisticated spatial arrangements \cite{Matusik:2005:TDU,Liu:2004:NRT}.

Recent deep learning-based methods employ parametric neural networks, which excel at representing textures, yet they often fall short in providing the interpretability and control crucial for various user editing operations \cite{Yu:2019:TM,Bergmann:2017:LTM,Henzler:2020:LN3,Hu:2024:DTP}.
Another line of active research focuses on transforming image textures into parametric, node-graph-based procedural textures \cite{Li:2023:EEP,Shi:2020:DMG,Guerrero:2022:MF,Hu:2023:GPM}.
\note{
These procedural textures, once created, offer a professional-grade level of control and quality over texture appearances.
}%
However, current methods are limited to a narrow range of texture categories and encounter difficulties in synthesizing graph structures \new{and faithfully reproducing the input image textures}.
\note{
Moreover, the accuracy of the generated graphs and their parameters falls short, failing to accurately and faithfully reproduce the appearance features of the input image textures.
}%
Therefore, existing approaches are either constrained to a limited set of texture editing capabilities, or their applicability tends to be confined to specific types of textures.
Our approach uses an autoencoder with a disentangled, compositional latent space, providing interpretability and extensive editing capabilities for various textures.

\paragraph{Compositional Neural Representations}
In certain data domains such as programming languages, vector graphics, \delete{social networks, }and engineering designs,
the underlying compositional structure is either inherently present or can be parsed using specific rules.
However, in other domains like 2D images and 3D reconstructions,
extracting and representing compositional structures remains a challenging task due to its elusive nature.
To address this, compositional neural representations have been proposed, utilizing a neural latent space to represent various data types compositionally.
This approach has been explored in several raster data types,
such as 2D images \cite{Epstein:2022:BGAN,Locatello:2020:OCL,Jiang:2023:OCS,Zhang:2018:UDO}, videos \cite{Elsayed:2022:SAVI++},
3D shapes \cite{Hertz:2022:EIS,Nash:2017:SVA}, and scenes \cite{Niemeyer:2021:GRS,Sajjadi:2022:OSR}.
These representations are derived using various methods,
ranging from unsupervised \cite{Epstein:2022:BGAN,Locatello:2020:OCL,Jiang:2023:OCS,Zhang:2018:UDO}\note{and}, semi-supervised \cite{Elsayed:2022:SAVI++}, to fully supervised methods \cite{Nash:2017:SVA}.
The structures of these neural representations include disentangled object slots \cite{Locatello:2020:OCL,Jiang:2023:OCS}, spatial points \cite{Zhang:2018:UDO}, graphs \cite{He:2022:AL,Ost:2021:NSG}, Gaussians \cite{Hertz:2022:EIS,Epstein:2022:BGAN}, and image segments \cite{He:2022:LSU}.
In this work, we introduce a compositional neural representation for image textures.
This representation is characterized by compositions of Gaussians within a disentangled neural latent space, and is learned without external supervision.

\section{Method}
\label{sec:method}

Given the conceptual and mathematical ambiguities surrounding textons, we outline 4 properties to characterize textons in Section~\ref{sec:method:properties}.
Section~\ref{sec:method:representation} proposes the compositional neural representation of textures, conceptualized through neural textons.
Section~\ref{sec:method:architecture} presents the autoencoder architecture used to learn the neural latent representation of these textons.
Section~\ref{sec:method:unsupervised} outlines unsupervised training strategies and losses for enforcing the desired properties of neural textons in the autoencoder's latent space.

\subsection{Properties of textons}
\label{sec:method:properties}

As texton is conceptually ambiguous, we determine 4 desired properties of neural textons which are essential for the textons to be interpretable and useful for texture editing.

\begin{textonprop}[Discrete Objectness]\label{property:discrete}
A texton embodies a distinct element within the image that can be identified as either present or absent in its entirety,
and a texture can be fully described by a discrete set of textons.
This discrete set representation facilitates intuitive and interpretable image editing, allowing users to engage in a more natural and user-friendly editing experience.
\end{textonprop}
\begin{textonprop}[Spatial Compactness]\label{property:compact}
A neural texton describes the spatial extent and appearance of a {\bf local} region within a texture.%
\end{textonprop}
\begin{textonprop}[Reshuffling Invarance]\label{property:reshuffle}
\replace{Textures inherently possess a stochastic characteristic that exhibits spatial stationarity.}{For a texture with a homogenous type/appearance, }%
random reshuffling of appearances among its textons should span the stochastic (sub)space of a texture without modifying the texture's overall appearance. That is, reshuffled texton appearances should result in a different version of the original texture.
\end{textonprop}
\begin{textonprop}[Transformation Equivariance]\label{property:transform}
When neural textons undergo a spatial transformation, the resulting image should reflect the same transformation, preserving the spatial relationships and appearance of the textons from the original texture.
\end{textonprop}

\new{These properties are enforced via a combined design of neural texton representation (Section~\ref{sec:method:representation}), autoencoder architecture (Section~\ref{sec:method:architecture}), and unsupervised training strategies (Section~\ref{sec:method:unsupervised}).}

\subsection{Textures as Compositions of Neural Textons} 
\label{sec:method:representation}

\replace{The encoder $\gaussianencoder$ maps an input texture $\image \in \realnumberset^{\imageheight\times\imagewidth\times 3}$ into our structured latent representation $\latent$,}{We represent an image texture as} a composition of Gaussians $\setof{\gaussian}{\gaussianindex}_{\gaussianindex=1}^{\numgaussians}$ \new{in the autoencoder latent space,}
where each Gaussian $\gaussian_{\gaussianindex}$ represents a texton and $\numgaussians$ is the\delete{(maximum)} number of Gaussians.
\new{The choice of Gaussian representation is motivated by its inherent compactness ({\scshape Property~\ref{property:compact}} ({\scshape Spatial Compactness})) and ability to intuitively approximate texton shape.}
To facilitate learning the optimal number of Gaussians required for a texture, we assign a weight parameter to each Gaussian $\gaussianweight \in [0, 1]$.
As we model textures as a discrete set of textons ({\scshape Property~\ref{property:discrete}} ({\scshape Discrete Objectness})),
$\gaussian_{\gaussianindex}$ exists (is valid) in the set if $\gaussianweight$ is near 1 and otherwise if $\gaussianweight$ is near 0. 
Each Gaussian is given by its center $\gaussianmean$ and covariance matrix $\gaussiancovariance$ that encodes the approximated shape of the texton.
To enhance the encoding of texton geometry and the subsequent geometry-appearance disentanglement, we include a 2D unit-vector $\gaussiandir \in \realnumberset^2$ ($\|\gaussiandir\|=1$) representing the anisotropic direction of textons in addition to Gaussian covariance.
\new{Using $\gaussiandir$ enables the network to more effectively model the directional properties of textons, such as those of an equilateral triangle shape, which are not fully captured by covariance alone.
}%
Furthermore, $\gaussian_\gaussianindex$ contains an appearance parameter given by a feature vector $\gaussianfeature \in \realnumberset^{\gaussianfeaturedim}$, which is used to encode local texton appearance.
\new{The separate encoding of texton geometry and appearance is fundamental to geometry-appearance disentanglement required by {\scshape Property~\ref{property:reshuffle}} ({\scshape Reshuffling Invarance}) and {\scshape Property~\ref{property:transform}} ({\scshape Transformation Equivariance}).}
\note{But Gaussian shape (center and covariance) explicitly encode directional information only when the Gaussian is anisotropic.}
In summary, our compositional Gaussian representation $\latent$ is defined as
\replace{ 
\begin{align}
\latent =\ & \setof{\gaussian}{\gaussianindex}_{\gaussianindex=1}^{\numgaussians} = \gaussianencoder(\image)
\end{align}
}
{
$\latent = \setof{\gaussian}{\gaussianindex}_{\gaussianindex=1}^{\numgaussians}$,
}
where 
\begin{math}
\gaussian_\gaussianindex=\{\gaussianweight, \gaussianmean,\gaussiancovariance,\gaussiandir,\gaussianfeature\}.
\label{eq:gaussian_properties}
\end{math}
Figure~\ref{fig:teaser:overview} visualizes the representation of 4 example textures as compositions of Gaussians.

\note{Figure~\ref{fig:method_architecture} shows an overview of our autoencoder and unsupervised training strategy, consisting of three major components: autoencoding, appearance reshuffling, and disentangling.
}%

\note{
In autoencoding (Figure~\ref{fig:method_architecture}  \begin{imageonly}$\autoencodingbox$\end{imageonly}), the encoder takes an input texture $\image$ and generates its latent Gaussians $\setof{\gaussian}{}$. Then, the decoder employs a Gaussian splatting layer $\gaussiansplatting$ to convert latent Gaussians into a feature map \cite{Epstein:2022:BGAN} that is fed into the generator $\gaussiangenerator$ to reconstruct the input texture $\imagerecon$.
However, autoencoding is not sufficient to reliably learn the extraction of textons as well as separation of texton geometry and appearance in a fully unsupervised manner.
}%
\note{
We then design the appearance reshuffling and disentangling components to maintain these properties.
}%

\note{
Our approach promotes {\scshape Property~\ref{property:discrete}} ({\scshape Discrete Objectness}) and {\scshape Property~\ref{property:compact}} ({\scshape Spatial Compactness}) by applying corresponding compactness $\losscompact$ and entropy $\lossentropy$ losses in the autoencoding part (Figure~\ref{fig:method_architecture}  
\begin{imageonly}$\autoencodingbox$\end{imageonly}). In addition to the losses, we employ the inherently compact Gaussians to represent textons. 
Aligned with {\scshape Property~\ref{property:reshuffle}} ({\scshape Reshuffling Invarance}), the reshuffling component modifies the Gaussian appearance features (Figure~\ref{fig:method_architecture} \begin{imageonly}$\reshufflingbox$\end{imageonly}).
These features are associated with input latent Gaussians $\setof{\gaussian}{}$, which are derived by encoding the input texture.
The generator $\gaussiangenerator$ generates a new image $\imagereshufflerecon$ from the reshuffled latent Gaussians $\setof{\gaussianreshuffle}{}$.
The generated image $\imagereshufflerecon$ is expected to closely resemble the input texture $\image$ in terms of texture similarity.
The disentangling component (Figure~\ref{fig:method_architecture} \begin{imageonly}$\disentanglingbox$\end{imageonly}) is to promote {\scshape Property~\ref{property:transform}} ({\scshape Transformation Equivariance}).
We transform the input texture and encode the tranformed texture $\imagetransformed$ as latent Gaussians $\setof{\gaussiantransformed}{}$.
The same transformation $\geometrictransform$ is applied directly to input latent Gaussians $\setof{\gaussian}{}$.
In accordance with {\scshape Property~\ref{property:transform}} ({\scshape Transformation Equivariance}), we expect that the Gaussians $\setof{\gaussiantransformed}{}$ derived from transformed input textures should closely match the transformed Gaussians $\geometrictransform(\setof{\gaussian}{})$ from the original input textures.
}%

\subsection{Network Architecture}
\label{sec:method:architecture}

The autoencoder (Figure~\ref{fig:teaser:overview}) has an encoder (Section~\ref{sec:method:encode}), a Gaussian splatting layer (Section~\ref{sec:gaussian_splatting}) and a generator (Section~\ref{sec:method:decode}).

\begin{figure}[tb]
\centering
    \includegraphics[height=0.9\linewidth]{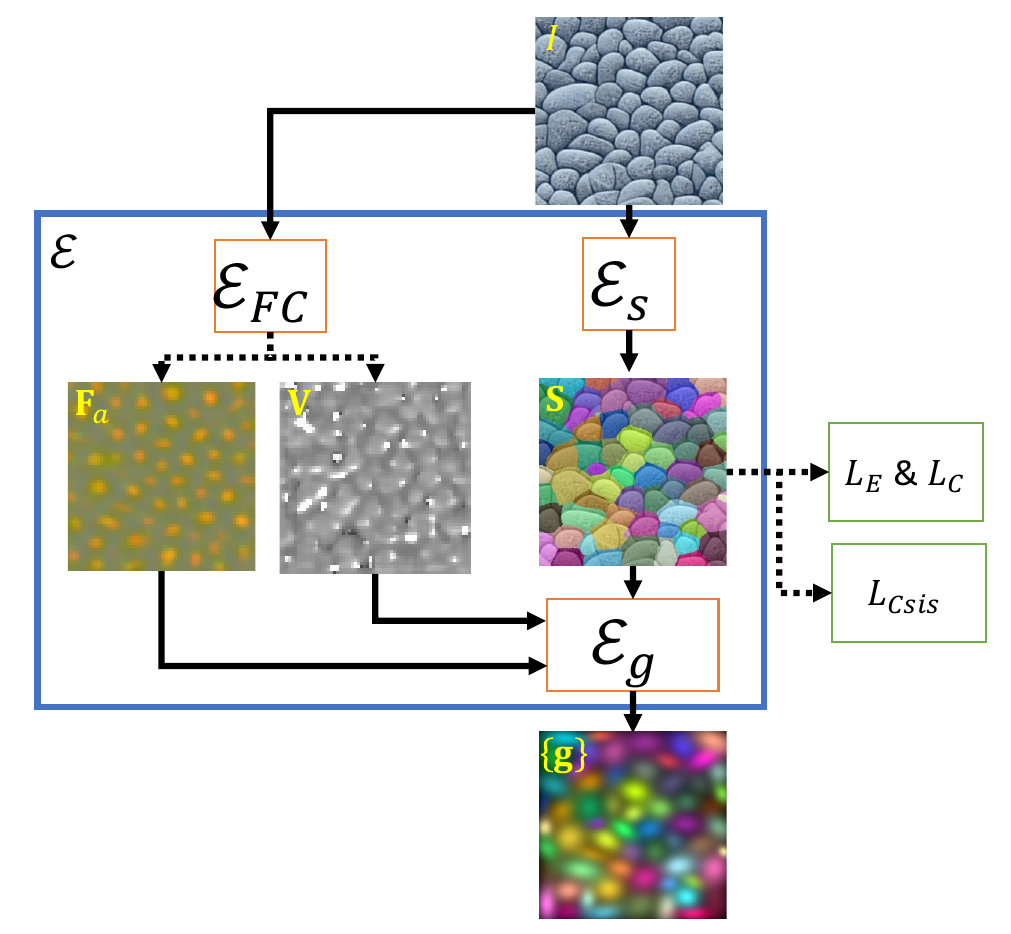}
\caption[]{\textit{Encoder mapping an image into latent Gaussians.}
    The encoder (Section~\ref{sec:method:encode}) comprises a FC image encoder $\fcencoder$, a segmentation network $\segmentationnetwork$, and a Gaussian parameter estimation layer $\gaussianestimationlayer$.
    $\fcencoder$ calculates an appearance feature map $\imageappearancefeature$ and direction map $\imagedirfeature$, which, along with segmentations $\segmap$ generated from $\segmentationnetwork$, are used to compute latent Gaussians $\setof{\gaussian}{}$.
    \replace{
    In accordance with {\scshape Property~\ref{property:discrete}} ({\scshape Discrete Objectness}) and  {\scshape Property~\ref{property:compact}} ({\scshape Spatial Compactness}), entropy $\lossentropy$ and compactness $\losscompact$ losses, are applied to $\segmap$ to foster hard and localized segmentation masks in order to inject these desired properties into textons.
    Segmentations $\segmap$ are also employed in computing $\lossmatch$, as detailed in Section~\ref{sec:method:unsupervised:disentangle}.
    }{Entropy $\lossentropy$, compactness $\losscompact$ and consistency $\lossmatch$ losses (Section~\ref{sec:method:unsupervised}) are applied to $\segmap$ to inject desired properties (Section~\ref{sec:method:properties}) into textons.\delete{, as discussed in Section~\ref{sec:method:unsupervised}.}}%
}
\label{fig:method_encoder}
\end{figure}

\subsubsection{Encoder}
\label{sec:method:encode}
\note{
To encode image $\image$ into latent Gaussians $\latent=\gaussiancomposition$, there are two options for generating the Gaussian parameters (\Cref{eq:gaussian_properties}): the first is to directly predict these parameters via linear or MLP heads; the second, which we choose, is to compute them indirectly from predicted heatmaps or segmentation maps.
The second option has demonstrated better performance in previous tasks such as human pose estimation and unsupervised landmark detection  \cite{Zheng:2023:DLB,Zhang:2018:UDO}.
Therefore,
}%
The encoder $\gaussianencoder$ includes an image encoder $\fcencoder$, a segmentation network $\segmentationnetwork$, and a Gaussian parameter estimation layer $\gaussianestimationlayer$.
\new{Figure~\ref{fig:method_encoder} illustrates the encoder architecture.}
\delete{The encoder }$\fcencoder$ is a shallow, fully convolutional (FC) network in order to avoid encoding non-local structure information, as per {\scshape Property~\ref{property:compact}} ({\scshape Spatial Compactness}). $\fcencoder$
takes input images $\image\in\realnumberset^{\imageheight\times\imagewidth\times 3}$ to generate feature maps, and has two branches: the appearance branch produces an appearance feature map $\imageappearancefeature \in \realnumberset^{\imageheight\times\imagewidth\times\imageappearancefeaturedim}$ and the geometry branch produces a direction map $\imagedirfeature \in \realnumberset^{\imageheight\times\imagewidth\times 2}$.
The Mask2Former architecture is used \cite{Cheng:2022:MAM} as $\segmentationnetwork$ and trained end-to-end with the other networks.
$\segmentationnetwork$ takes the image $\image$ and generates pixel-wise, softmax-normalized $\numgaussians$ segmentation masks $\segmap  \in [0, 1]^{\numgaussians\times\imageheight\times\imagewidth}$. \new{Each segment is used to compute a Gaussian. $\numgaussians$ is fixed and is also the number of Gaussians}.
For more details about $\fcencoder$ and $\segmentationnetwork$, please refer to Appendix D.1.
Given the $\numgaussians$-segmentation masks $\segmap$, the appearance feature map $\imageappearancefeature$, and the directional map $\imagedirfeature$, $\gaussianestimationlayer$ computes the composition of Gaussians $\gaussiancomposition$, as detailed below.
\paragraph{Spatial parameters}
The mean $\gaussianmean=\frac{\sum_{\coordvec}\coordvec \cdot \segment(\coordvec)}{\normsegmap}$ and covariance $\gaussiancovariance= \frac{\sum_{\coordvec}(\coordvec - \gaussianmean)^T (\coordvec - \gaussianmean) \cdot \segment(\coordvec)}{\normsegmap}$ of the $\gaussianindex^{\text{th}}$ Gaussian $\gaussian_{\gaussianindex}$ is computed as the average and covariance of image coordinates $\coordvec=[\xcoord,\ycoord]$ weighted by the $\gaussianindex^{\text{th}}$ segmentation mask $\segmap_{\gaussianindex}$,
\delete{
\begin{align}
\gaussianmean =&  \frac{\sum_{\coordvec}\coordvec \cdot \segment(\coordvec)}{\normsegmap} \\
\gaussiancovariance =&  \frac{\sum_{\coordvec}(\coordvec - \gaussianmean)^T (\coordvec - \gaussianmean) \cdot \segment(\coordvec)}{\normsegmap}
\end{align}
}%
where $\normsegmap=\sum_{\coordvec}\segment(\coordvec)$ is the normalization factor computed from the ${\gaussianindex}^{\text{th}}$ segment.
The direction $\gaussiandir$ is computed as the weighted average (pooling) of directional map $\imagedirfeature$ within each image segment as
\begin{equation}
\gaussiandir= \frac{\sum_{\coordvec}\imagedirfeature(\coordvec) \segment(\coordvec)}{\normsegmap},
\label{eqn:gaussian_direction}
\end{equation}
followed by normalization $\gaussiandir=\gaussiandir/\|\gaussiandir\|$.

\paragraph{Gaussian weight} The Gaussian weight $\gaussianweight$ is used to adaptively adjust the number of \emacsquote{valid} Gaussians by indicating the existence of each Gaussian. While this existence weight should ideally be binary as indicated by {\scshape Property~\ref{property:discrete}} ({\scshape Discrete Objectness}), during model training it is crucial to ensure that gradients are effectively back-propagated through the entire network.
Hence, we introduce a differentiable, randomized sampling mechanism for computing $\gaussianweight \sim \bernoullidistribution(\bernoulliprob_\gaussianindex)$ as a sample from Bernoulli distribution $\bernoullidistribution$. $\gaussianweight$ takes the value 1 with the probability $\bernoulliprob_\gaussianindex= \frac{\sum_{\coordvec} \segment(\coordvec) \segment(\coordvec)}{\normsegmap}$ computed
\delete{
as
\begin{equation}
\bernoulliprob_{\gaussianindex}= \frac{\sum_{\coordvec} \segment(\coordvec) \segment(\coordvec)}{\normsegmap}.
\label{eq:bernoulli_prob}
\end{equation}
\Cref{eq:bernoulli_prob} computes $\bernoulliprob_{\gaussianindex}$}%
as the average of mask $\segmap_{\gaussianindex}$ weighted by $\segmap_{\gaussianindex}$.
\note{
By computing the average of these mask probabilities weighted by the probabilities themselves, we essentially emphasize the higher probabilities, as higher values will contribute more to the average.
}%
If the mask has either high probability values (such as in a local region) or zero values (like those outside the region), which might indicate a generally reliable mask, $\bernoulliprob_{\gaussianindex}$ will be high and $\gaussianweight \sim \bernoullidistribution(\bernoulliprob_\gaussianindex)$, thus the Gaussian $\gaussian_\gaussianindex$ is more likely to exist ($\gaussianweight=1$).
Conversely, if many pixels have low but non-zero probability values, implying uncertainty in the segmentation, $\bernoulliprob_{\gaussianindex}$ will be low and the Gaussian $\gaussian_\gaussianindex$ is more likely not to exist ($\gaussianweight=0$).
Because sampling from a (categorical) Bernoulli distribution is not differentiable, during training we apply the Gumbel-softmax trick \cite{Jang:2016:CRG}  to approximate the hard sampling process from a categorical distribution.
During inference, we round the continuous values of $\bernoulliprob_{\gaussianindex} \in [0, 1]$ to obtain binary values $\gaussianweight \in \{0,1\}$, aligning with {\scshape Property~\ref{property:discrete}} ({\scshape Discrete Objectness}).

We find that using the categorical sampling mechanism ($\gaussianweight \sim \bernoullidistribution(\bernoulliprob_\gaussianindex)$) \replace{when computing Gaussian weights}{during training} generates more aligned and more reliable texton segmentations and thus higher-quality latent Gaussians than directly setting $\gaussianweight$ as $\bernoulliprob_{\gaussianindex}$.
We hypothesize that the sampling mechanism inherently promotes weights to converge towards 0 or 1, thereby ensuring binary outcomes ({\scshape Property~\ref{property:discrete}} ({\scshape Discrete Objectness})),
while simultaneously maintaining {\it exploration} capabilities through randomized sampling when searching for good texton representations.
\new{We could not achieve the same balanced effects by adjusting the training losses or loss schedulers without this sampling mechanism.}

\paragraph{Appearance parameters}
Similarily to Equation~(\ref{eqn:gaussian_direction}), the appearance feature $\gaussianfeature= \frac{\sum_{\coordvec}\imageappearancefeature(\coordvec) \segment(\coordvec)}{\normsegmap}$
\delete{$\gaussiandir= \frac{\sum_{\coordvec}\imagedirfeature(\coordvec) \segment(\coordvec)}{\normsegmap}$}\replace{and direction $\gaussiandir$ are}{is}\delete{computed as} the weighted average of image appearance features $\imageappearancefeature$, \delete{and directional map $\imagedirfeature$ }\note{within each image segment as}
\delete{$\gaussiandir= \frac{\sum_{\coordvec}\imagedirfeature(\coordvec) \segment(\coordvec)}{\normsegmap}$}%
followed by normalization $\gaussianfeature=\gaussianfeature/\|\gaussianfeature\|$.
\delete{$\gaussiandir=\gaussiandir/\|\gaussiandir\|$.}

\subsubsection{Gaussian Splatting}
\label{sec:gaussian_splatting}
After obtaining latent Gaussians $\gaussiancomposition$, the Gaussian splatting layer converts $\gaussiancomposition$ into a feature grid $\gaussiansplattedfeature$, 
which is then fed into a generator to reconstruct the input image as $\imagerecon$.
The conversion follows the implementation of Gaussian splatting in \cite{Epstein:2022:BGAN} with minor modifications and is visualized as $\gaussiansplatting$ in Figure~\ref{fig:method_architecture}. 
For more details, please see Appendix D.1.

\note{
}%

\subsubsection{Generator}
\label{sec:method:decode}

We adopt the SPADE generator \cite{Park:2019:SIS} as the generator $\gaussiangenerator$  to reconstruct $\imagerecon$ from $\gaussiansplattedfeature$. 
The original SPADE generator takes a learned constant feature map and is injected with resized $\gaussiansplattedfeature$ at multiple levels, which results in contant output sizes during inference.
We replace the learned constant feature map with feature maps from $\gaussiansplatting$ so $\gaussiangenerator$ can generate textures of arbitrary resolutions after training.
\note{
Instead, we replace the learned constant feature map with downscaled $\generatorinputfeature=\max_{\gaussianindex} \splatalpha_{\gaussianindex}$
 as input to SPADE so it can generate textures of arbitrary resolutions.
}%
For more details, please refer to Appendix D.1.

\subsubsection{Image reconstruction}
The reconstruction is ensured using a mixed reconstruction loss function
\begin{equation}
\lossrecon = \weightreconnorm\lossreconnorm +  \weightlpips \losslpips
\label{eq:recon}
\end{equation}
where $ \lossreconnorm=\|\image-\imagerecon\|_1$ and $ \losslpips$ denotes LPIPS loss \cite{Zhang:2018:UED} measuring perceptual difference between $\image$ and $\imagerecon$.
We set $\weightreconnorm=2$ and $\weightlpips=0.2$.
In addition, we wish for the reconstructed images $\imagerecon$ to be realistic by employing a discriminator $\discriminator$. We apply the non-saturating adversarial loss on the reconstructed images to train the encoder $\gaussianencoder$ and generator $\gaussiangenerator$
\begin{equation}
\lossgan= -\log\left(\discriminator\left(\imagerecon\right)\right).
\end{equation}
For details about the discriminator $\discriminator$ and its losses, see Appendix D. Figure~\ref{fig:method_architecture}  \begin{imageonly}$\autoencodingbox$\end{imageonly} shows the process of encoding ($\gaussianencoder$) image as Gaussians and decoding ($\gaussiansplatting$, $\gaussiangenerator$) Gaussians as the original image.

\subsection{Unsupervised Learning of Neural Textons}
\label{sec:method:unsupervised}

\note{We utilize Gaussians to represent textons which we wish for exhibiting four properties, as discussed in \replace{Section~\ref{sec:method:overview}}{Section~\ref{sec:method:encode}}.
}%
In this section, we introduce unsupervised learning strategies to inject the 4 properties (Section~\ref{sec:method:properties}) into the neural textons.

\begin{figure}[tb]
\centering
    \includegraphics[height=0.85\linewidth]{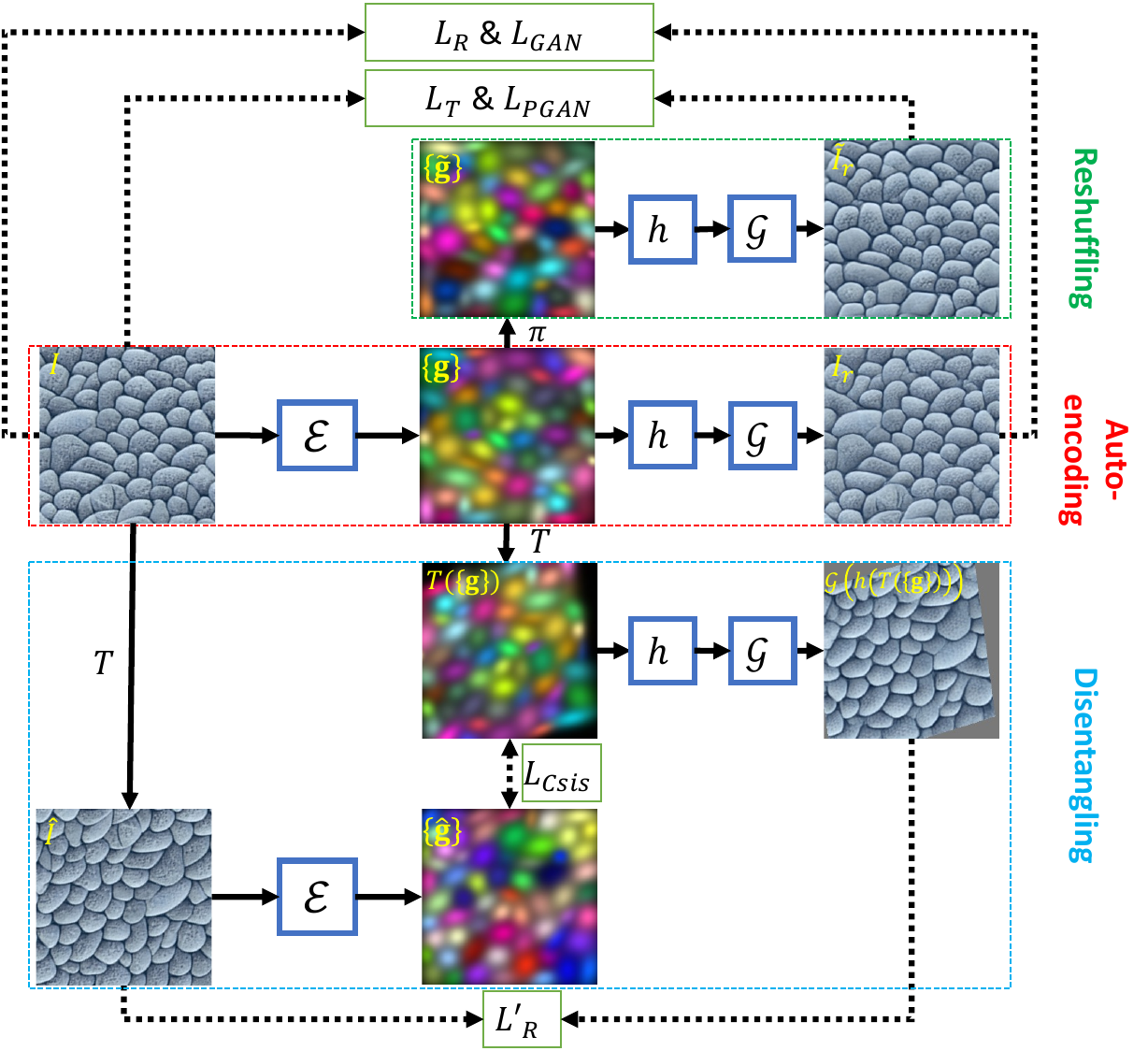}
\caption[]{\textit{Architecture of the proposed method.}
    Our network is composed of three branches:  \begin{imageonly}$\autoencodingbox$\end{imageonly} (Sections~\ref{sec:method:encode} to \ref{sec:method:decode}), texton \begin{imageonly}$\reshufflingbox$\end{imageonly} and structure-appearance \begin{imageonly}$\disentanglingbox$\end{imageonly} (Section~\ref{sec:method:unsupervised}).%
}
\label{fig:method_architecture}
\end{figure}

\textbf{{\scshape Property~\ref{property:discrete}} ({\scshape Discrete Objectness})}
In the process of encoding textures as Gaussisans $\gaussiancomposition$, we calculate several losses on the intermediate segmentation masks $\segmap$ during end-to-end training (Figure~\ref{fig:method_encoder}).
One is the entropy loss $\lossentropy$, which encourages reliable hard segmentation\note{$\segmap$} and thus discrete textons with $\gaussianexistence \rightarrow 1$ or $\gaussianexistence \rightarrow 0$\delete{(\Cref{eq:bernoulli_prob})},
\begin{equation}
\lossentropy=- \frac{1}{\imageheight\imagewidth}\sum_{\coordvec}  \sum_{i=1}^{\numgaussians} \segment(\coordvec) \log(\segment(\coordvec)).
\end{equation}
$\lossentropy$ is minimized when segmentation masks $\segmap$ are hard with probabilities either 0 or 1, or equivalently when $\gaussianexistence=0$ or $1$.

\textbf{{\scshape Property~\ref{property:compact}} ({\scshape Spatial Compactness})}
We also apply a compactness loss $\losscompact$ \cite{Yang:2020:SSF}, especially during the initial phase of training (see Appendix D.3) on the segmentation masks so that each segment $\segmap_\gaussianindex$ and thus $\gaussian_\gaussianindex$ is spatially compact and localized,
\begin{align}
\losscompact= \frac{1}{\imageheight\imagewidth} \sum_{\coordvec}\norm{\reconcoordvec  - \coordvec}_2^2,
\end{align}
where $\reconcoordvec = \sum_{\gaussianindex} \gaussianmean \cdot \segment(\coordvec)$.

\textbf{{\scshape Property~\ref{property:reshuffle}} ({\scshape Reshuffling Invarance})}
\note{
Complex reshuffling invariance; why is it complex; reasons; why there is no easy solutions; interesting
}%
\label{sec:method:unsupervised:reshuffle}
\note{In accordance with {\scshape Property~\ref{property:reshuffle}} ({\scshape Reshuffling Invarance}), }Random reshuffling of texton appearances $\gaussianfeaturesymbol$ (with other geometric parameters unchanged) should result in a different version of original texture without altering overall texture appearance.
More specifically, {\it appearance feature reshuffling}, denoted as $\reshufflesymbol$, permutes appearance features within Gaussians to obtain another set of Gaussians $\setof{\gaussianreshuffle}{\gaussianindex}=\reshufflesymbol(\setof{\gaussian}{\gaussianindex})$.
In this randomized process $\gaussiancompositionreshuffle=\reshufflesymbol(\gaussiancomposition)$,
when the appearance feature $\gaussianfeature$ in $\gaussian_\gaussianindex$ moves to $\gaussian_\gaussianindexj$ originally with feature $\gaussianfeaturesymbol_{\gaussianindexj}$,
$\gaussian_\gaussianindexj$ becomes $\gaussianreshuffle_\gaussianindexj$ and the resulting appearance feature $\tilde{\gaussianfeaturesymbolfunc{\gaussianindexj}}$ in $\gaussianreshuffle_\gaussianindexj$ is calculated as
\begin{equation}
\tilde{\gaussianfeaturesymbolfunc{\gaussianindexj}} =   \reshufflefactor\gaussianfeaturesymbolfunc{\gaussianindex} + (1-\reshufflefactor) \gaussianfeaturesymbolfunc{\gaussianindexj}.
\label{eq:reshuffling}
\end{equation}
where $\reshufflefactor \in [0, 1]$ is the reshuffling coefficient from $\gaussianfeaturesymbolfunc{\gaussianindex}$ to $\gaussianfeaturesymbolfunc{\gaussianindexj}$.
During inference, the {\it appearance feature reshuffling} is only applied among valid discrete textons with their Gaussian weights $\gaussianweightsymbol$ rounded as 1 (while invalid ones have $\gaussianweightsymbol=0$), and $\reshufflefactor$ is computed as
\replace{
Formally, during inference if $\gaussianweight=1$ {\em and} $\gaussianweightsymbol_{\gaussianindexj}=1$, $\reshufflefactor=1$ $\Rightarrow \tilde{\gaussianfeaturesymbolfunc{\gaussianindexj}}=\gaussianfeaturesymbolfunc{\gaussianindex}$;
otherwise if $\gaussianweight=0$ {\em or} $\gaussianweightsymbol_{\gaussianindexj}=0$, $\reshufflefactor=0$ $\Rightarrow \tilde{\gaussianfeaturesymbolfunc{\gaussianindexj}}=\gaussianfeaturesymbolfunc{\gaussianindexj}$.
}{%
\begin{align}
\reshufflefactor = \gaussianweight \gaussianweightsymbol_{\gaussianindexj}
\end{align} 
}%
\note{we can describe the reshuffling process \new{within valid textons ($\gaussianweightsymbol=1$)} with a \note{square binary}permutation matrix.}%
In the training phase, however, as the network evolves to determine the ideal number of Gaussians representing textons by continuously adjusting the Gaussian weight $\gaussianweightsymbol$,
it is crucial to \emacsquote{prioritize} {\it appearance feature reshuffling} among Gaussians that are more likely to represent \note{individual }textons.
While $\gaussianweight$ can be used to determine the likelihood of a Gaussian representing a texton after effective network training, $\gaussianweight$ is not a reliable indicator of likelihood due to the immaturity of the texton representation during training.
For example, even if $\gaussianweight$ is close to 1, the Gaussian may not effectively represent a texton if the area of its corresponding texton segmentation mask $\segmentarea=\sum_{\coordvec}\segment$ is close to 0.
Therefore, we define the likelihood score of a Gaussian $\gaussian_\gaussianindex$ representing a texton as a tuple $(\segmentarea,\gaussianexistence)$.
A Gaussian $\gaussian_\gaussianindex$ is either likely (if $\segmentareasymbol_{\gaussianindex} \gg 0$ and $\gaussianexistencesymbol_{\gaussianindex} \gg 0$) or unlikely (if $\segmentareasymbol_{\gaussianindex} \approx 0$ or $\gaussianexistencesymbol_{\gaussianindex} \approx 0$) to represent a texton.
To prioritize reshuffling among Gaussians that are more likely to represent textons, the reshuffling coefficient $\reshufflefactor$ is computed as
\begin{equation}
\reshufflefactor=\left(\min\left(\max\left(\frac{\segmentarea}{\segmentareasymbol_{\gaussianindexj}},\frac{\gaussianexistence}{\gaussianexistencesymbol_{\gaussianindexj}}\right),1\right)\right)^{\reshufflepower}.
\label{eq:reshuffling_factor}
\end{equation}
\replace{
In Equation~(\ref{eq:reshuffling_factor}), there are 4 cases as follows:
1) Both Gaussians are likely to represent textons and eshuffling is not ignored from $\gaussian_\gaussianindex$ to $\gaussian_\gaussianindexj$ with $ 0 \ll \reshufflefactor < 1$.
2) $\gaussiansymbol_\gaussianindex$ but $\gaussiansymbol_\gaussianindexj$ is likely to represent a texton, reshuffling is ignored with $\reshufflefactor\approx 0$.
3) If $\gaussian_\gaussianindexj$ but $\gaussian_\gaussianindex$ is likely to represent a texton; reshuffling is also ignored even with $\reshufflefactor \in [0, 1]$, since the Gaussian splatting layer $\gaussiansplatting$ will ignore both $\gaussiansymbol_\gaussianindex$ and $\gaussiansymbol_\gaussianindexj$ when their areas ($\segmentareasymbol$) or weights ($\gaussianweightsymbol$) approach zero.
4) If neither Gaussian is likely to represent a texton.
}%
{%
Intuitively, Equation~(\ref{eq:reshuffling_factor}) will produce a higher value of $\reshufflefactor$ if $\gaussian_\gaussianindex$ is more likely than $\gaussian_\gaussianindexj$ to represent a texton.
And for the case where $\segmentareasymbol_{\gaussianindexj} \approx 0$ or $\gaussianexistencesymbol_{\gaussianindexj} \approx 0$, $\gaussian_\gaussianindexj$ will be ignored during the Gaussian splatting process regardless of its $\reshufflefactor$ value.
}
$\reshufflepower=0.5$ is a non-negative temperature hyperparameter controls the softness of the reshuffling operation, which will be discussed in Appendix D.2.

\note{
For example, when the Gaussian weight/existence $\gaussianweight/\gaussianexistence$ or the area of its corresponding texton segmentation mask $\segmentarea=\sum_{\coordvec}\segment$ is very small or approaching 0  ($\gaussianexistence\approx 0$ or $\gaussianweight\approx 0$ or $\segmentarea\approx 0$),
we anticipate that the latent Gaussian $\gaussian_\gaussianindex$ and its associated appearance feature $\gaussianfeature$ do not effectively represent a texton and should, therefore, be \replace{(basically) excluded}{discouraged} from appearance feature reshuffling.
}%
\replace{we randomly reshuffle, denoted as $\reshufflesymbol$, appearance feature $\gaussianfeature$ associated with $\gaussian_\gaussianindex$ within $\setof{\gaussian}{\gaussianindex}$ to obtain $\setof{\gaussianreshuffle}{\gaussianindex}=\reshufflesymbol(\setof{\gaussian}{\gaussianindex})$, which is then reconstructed}{The Gaussians $\setof{\gaussianreshuffle}{\gaussianindex}$ with reshuffled features are reconstructed} as the image $\imagereshufflerecon=\gaussiangenerator\left(\gaussiansplatting(\setof{\gaussianreshuffle}{\gaussianindex})\right)$ that should resemble the input image $\image$ \new{in terms of texture similarity}.
\note{In accordance with {\scshape Property~\ref{property:reshuffle}} ({\scshape Reshuffling Invarance}),}%
\note{Appearance feature reshuffling should not alter overall texture appearance.}%
To this end, we calculate texture loss $\losstexture$ between the original input $\image$ and image $\imagereshufflerecon$ by using the patch-based texture loss \note{defined in} \cite{Zhou:2023:NTS}
and non-saturating GAN losses on image patches \new{$\losspatchgan$} \cite{Park:2020:SAD} to train $\gaussianencoder$ and $\gaussiangenerator$.
Please refer to Appendix D.2 for more implementation details.
\new{While enforcing {\scshape Property~\ref{property:compact}} ({\scshape Spatial Compactness}) encourages small segmentations/Gaussians, enforcing {\scshape Property~\ref{property:reshuffle}} ({\scshape Reshuffling Invarance}) encourages segmentations/Gaussians to capture the shape and location of individual textons so that reshuffling won't destory texture structure.
}%
Figure~\ref{fig:method_architecture} \begin{imageonly}$\reshufflingbox$\end{imageonly} shows the reshuffling operation $\reshufflesymbol$ and losses to enforce {\scshape Property~\ref{property:reshuffle}} ({\scshape Reshuffling Invarance}).

\note{\Cref{eq:reshuffling} \delete{and \Cref{eq:reshuffling_factor}}define the basic pairwise feature reshuffling operation, namely feature swapping, used in the feature reshuffling $\reshufflesymbol$ process.}%

\note{
Since $\gaussianweight$ and $\gaussianexistence$ are similar concepts, we find that using one of them $\gaussianexistence$ is sufficient for computing $\reshufflefactor$ in \Cref{eq:reshuffling_factor}.
A Gaussian can either effectively represent a neural texton or not.
We consider a Gaussian $\gaussian$ to effectively represent a texton if both associated $\segmentareasymbol \gg 0$ and $\gaussianexistencesymbol \gg 0$.
Conversely, a Gaussian $\gaussian$ is deemed ineffective in representing a texton if $\segmentareasymbol \approx 0$ or $\gaussianexistencesymbol \approx 0$.
where
}%

\note{
Accordingly, there are four distinct cases to consider when computing the reshuffling coefficient $\reshufflefactor$ (\Cref{eq:reshuffling}) between a pair of Gaussians $\gaussian_\gaussianindex, \gaussian_\gaussianindexj$, which are detailed below.
\begin{enumerate}
\item If both $\gaussian_\gaussianindex$ ($\segmentareasymbol_\gaussianindex \gg 0$ and $ \gaussianexistencesymbol_\gaussianindex \gg 0$) and $\gaussian_\gaussianindexj$  ($\segmentareasymbol_\gaussianindexj \gg 0$ and $ \gaussianexistencesymbol_\gaussianindexj \gg 0$) are effective in representing textons, the swapping is effective with $\reshufflefactor \gg 0$.

\item If $\gaussian_\gaussianindex$ is effective ($\segmentareasymbol_\gaussianindex \gg 0$ and $ \gaussianexistencesymbol_\gaussianindex \gg 0$), but $\gaussian_\gaussianindexj$ is not ($\segmentareasymbol_\gaussianindexj \approx 0$ or $ \gaussianexistencesymbol_\gaussianindexj \approx 0$), Equation~(\ref{eq:reshuffling_factor}) gives us $\reshufflefactor = 1$ and thus the swapping is effective. 
But if $\gaussianexistencereshuffle=\gaussianexistencesymbol_\gaussianindexj \approx 0$ ($\gaussianexistencereshuffle \in \gaussianreshuffle_\gaussianindexj$), the swapping has no effect on the reconstructed images and is in fact ineffective, because the Gaussian splatting (Section~\ref{sec:gaussian_splatting}) can ignore Gaussians with $\gaussianexistencereshuffle\approx 0$ (and thus $\gaussianweightreshuffle \approx 0$).

\item If $\gaussian_\gaussianindex$ is not effective ($\segmentareasymbol_\gaussianindex \approx 0$ or $ \gaussianexistencesymbol_\gaussianindex \approx 0$), but $\gaussian_\gaussianindexj$ is effective ($\segmentareasymbol_\gaussianindexj \gg 0$ and $ \gaussianexistencesymbol_\gaussianindexj \gg 0$), there are three subcases: the swapping is not effective with $\reshufflefactor \approx 0$ when both $\segmentareasymbol_\gaussianindex \approx 0$ and $ \gaussianexistencesymbol_\gaussianindex \approx 0$, the swapping is effective with $\reshufflefactor \gg 0$ but has no effect on the reconstruction when $\segmentareasymbol_\gaussianindex \gg 0$ and $ \gaussianexistencesymbol_\gaussianindex \approx 0$, and otherwise the swapping is effective.

\item If neither $\gaussian_\gaussianindex$ ($\segmentareasymbol_\gaussianindex \approx 0$ or $ \gaussianexistencesymbol_\gaussianindex \approx 0$)  nor $\gaussian_\gaussianindexj$ ($\segmentareasymbol_\gaussianindexj \approx 0$ or $ \gaussianexistencesymbol_\gaussianindexj \approx 0$) are effective, $\reshufflefactor$ can be any values between $[0, 1]$.
Even if it is possible that $\reshufflefactor \gg 0$, the swapping may still be ineffective   if $\gaussianexistencereshuffle=\gaussianexistencesymbol_\gaussianindexj \approx 0$ and thus $\gaussianweightreshuffle \approx 0$.
\end{enumerate}

}%

\replace{\subsubsection{Transformation Equivariance}
}
{
\textbf{{\scshape Property~\ref{property:transform}} ({\scshape Transformation Equivariance})}
}
\label{sec:method:unsupervised:disentangle}
\note{our objective is to achieve consistent results by first extracting the Gaussians and then applying transformations,
as compared to first transforming the image and subsequently transforming the Gaussians.
The consistency guarantees that the latent Gaussians
effectively separate the encoding of texton geometry and appearance, and maintain robustness against transformations applied to the image.
Concretely,}%
A consistency loss $\lossmatch$ is defined to reduce the differences between two sets of Gaussians $\geometrictransform(\gaussianencoder(\image))$ and $\gaussianencoder(\geometrictransform(\image))$ with associated segmentation masks.
The first set, $\geometrictransform(\gaussianencoder(\image))$, is derived by initially extracting Gaussians from the input texture $\image$ using our encoder $\gaussianencoder$, and subsequently applying the transformation $\geometrictransform$;
the second set, $\gaussianencoder(\geometrictransform(\image))$, is obtained by first applying $\geometrictransform$ to the input texture $\image$, and then extracting Gaussians with the encoder $\gaussianencoder$.
Similarily to \cite{Cheng:2022:MAM,Carion:2020:EOD}, we compute the bipartite matching $\matching$ between the two sets of Gaussians and segmentation masks and $\lossmatch$ is the sum of differences between matched Gaussians and masks.
In addition, we calculate the reconstruction loss $\lossreconprime$, using the same formulation as in Equation~(\ref{eq:recon}), but between transformed images $\imagetransformed=\geometrictransform(\image)$ and images reconstructed from $\geometrictransform\left(\gaussianencoder(\image)\right)$ while masking out regions transformed outside image boundaries.
Please refer to Appendix D.2 for more details.
Figure~\ref{fig:method_architecture} \begin{imageonly}$\disentanglingbox$\end{imageonly} shows the training mechanism to enforce {\scshape Property~\ref{property:transform}} ({\scshape Transformation Equivariance}).

\subsection{Overall Losses}

\replace{To sum up, we train the generation network ($\gaussianencoder, \gaussiansplatting$ and  $\gaussiangenerator$) with the following losses
\begin{equation}
\begin{split}
\loss = \lossrecon + \lossreconprime + \weightentropy\lossentropy  + 
\weightcompact\losscompact +  \weightmatch\lossmatch +  \weighttexture\losstexture + \\ \weightgan\lossgan +\weightpatchgan\losspatchgan
\end{split}
\end{equation}
}{Our final objective function for the encoder $\gaussianencoder$ and generator $\gaussiangenerator$ is $\loss = \lossrecon + \lossreconprime + \weightentropy\lossentropy  + 
\weightcompact\losscompact +  \weightmatch\lossmatch +  \weighttexture\losstexture + \weightgan\lossgan +\weightpatchgan\losspatchgan$.}
During training, we fix $\weighttexture, \weightgan$ and $\weightpatchgan$.
\replace{We adjust $\weightentropy, \weightcompact, \weightmatch$ during training to maintain a balance in the unsupervised training process.
This ensures that the reduction of one loss does not disproportionately occur at the expense of the others.}{While $\weightentropy, \weightcompact$, $\weightmatch$ are adjused to maintain a balance in the unsupervised training process.}
See Appendix D.3 for more training details.

\subsection{Dataset}
\label{sec:method:data}

We collect a high-quality texture dataset comprising 377K synthetic and 1.2K real stock images.
The synthetic data is from \replace{a commerical text-to-image diffusion model}{Adobe Firefly's text-to-image model} using 2500 detailed texture descriptions generated with \replace{a large language model (LLM)}{ChatGPT}.
We use 90\% images for training and 10\% images for testing (applications and analysis).
See Appendix A for some random data samples, data creation and augmentation details.

\begin{figure}[htb]
	\centering
	\subfloat[Input]{%
	\label{fig:diverse_texture_synthesis_main:exemplar} %
	\includegraphics[width=0.24\linewidth]{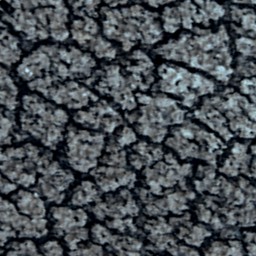}
	}%
	\subfloat[Output 1]{%
	\label{fig:diverse_texture_synthesis_main:output1}%
	\includegraphics[width=0.24\linewidth]{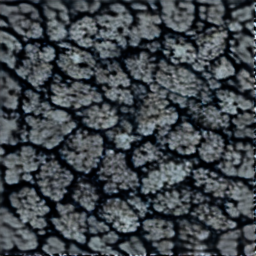}
	}%
	\subfloat[Output 2]{%
	\label{fig:diverse_texture_synthesis_main:output2}%
	\includegraphics[width=0.24\linewidth]{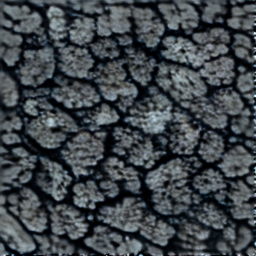}
	}%
	\subfloat[Output 3]{%
	\label{fig:diverse_texture_synthesis_main:output3}%
	\includegraphics[width=0.24\linewidth]{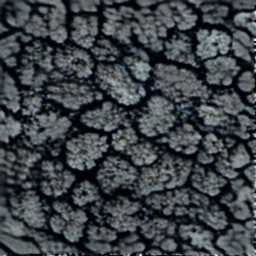}
	}%
	\caption[]{\textit{Texture diversification.}
		Given the input (a), 
		we can generate different versions
		(b) (c) (d) of the same texture by randomly reshuffling appearance features in latent Gaussians.
		(a) $\copyright$Tandem Stock (stock.adobe.com).

	}%
	\label{fig:diverse_texture_synthesis_main}
\end{figure}

\begin{figure}[htb]
	\centering
	\captionsetup[subfigure]{labelformat=empty}

    \subfloat[]{%
    \includegraphics[width=0.25\linewidth]{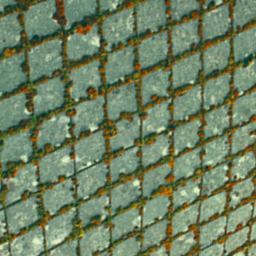}
    }%
    \subfloat[]{%
    \includegraphics[width=0.25\linewidth]{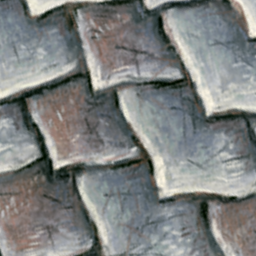}
    }%
    \subfloat[]{%
    \includegraphics[width=0.25\linewidth]{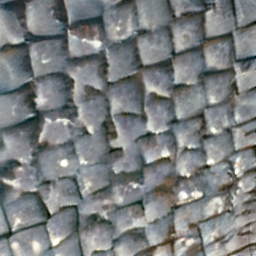}
    }%
    \subfloat[]{%
    \includegraphics[width=0.25\linewidth]{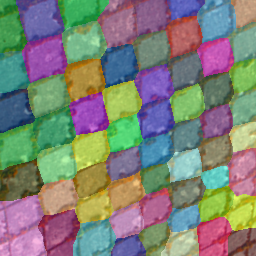}
    }%

    \vspace{-0.8cm}%
    \setcounter{subfigure}{0} %
	\captionsetup[subfigure]{labelformat=parens}
    \subfloat[Structure]{%
    \includegraphics[width=0.25\linewidth]{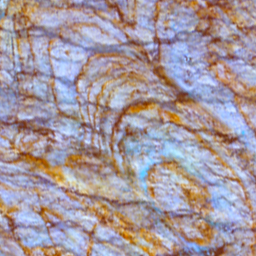}
    \label{fig:qualitative_texture_transfer:structure}%
    }%
    \subfloat[Appearance]{%
    \includegraphics[width=0.25\linewidth]{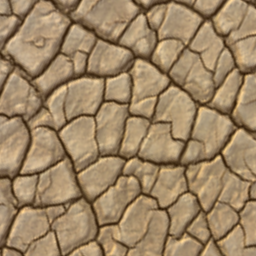}
    \label{fig:qualitative_texture_transfer:appearance}%
    }%
    \subfloat[Output]{%
    \includegraphics[width=0.25\linewidth]{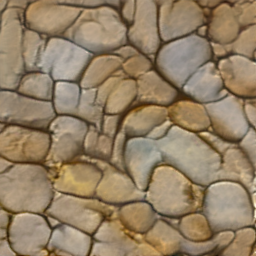}
    \label{fig:qualitative_texture_transfer:output}%
    }%
    \subfloat[Seg. overlay]{%
    \includegraphics[width=0.25\linewidth]{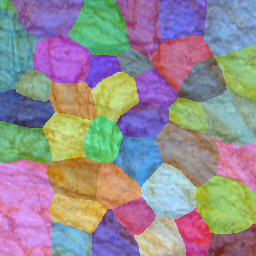}
    \label{fig:qualitative_texture_transfer:structure_overlay}%
    }%

\caption[]{\textit{Texture transfer.} Given structure-providing (a) and appearance-providing (b) textures, the output (c) combines the structure from one texture and appearance from the other.
We also visualize the segmentation masks (d) of the structure-providing inputs.
(a): (top) $\copyright$Westend61, (bottom) $\copyright$ADDICTIVE STOCK;
(b): (top) $\copyright$Liliya Rodnikova/Stocksy, (bottom) $\copyright$Elena Saurius\&Dani Rex/Stocksy.
}
\label{fig:qualitative_texture_transfer}
\end{figure}

\begin{figure}[h]
	\centering

	\setlength{\fboxrule}{2pt} %
	\setlength{\fboxsep}{0pt} %

	\subfloat[Plain transfer]{%
	\begin{tikzpicture}%
	\node (base) at (0,0) {\includegraphics[width=0.25\linewidth]{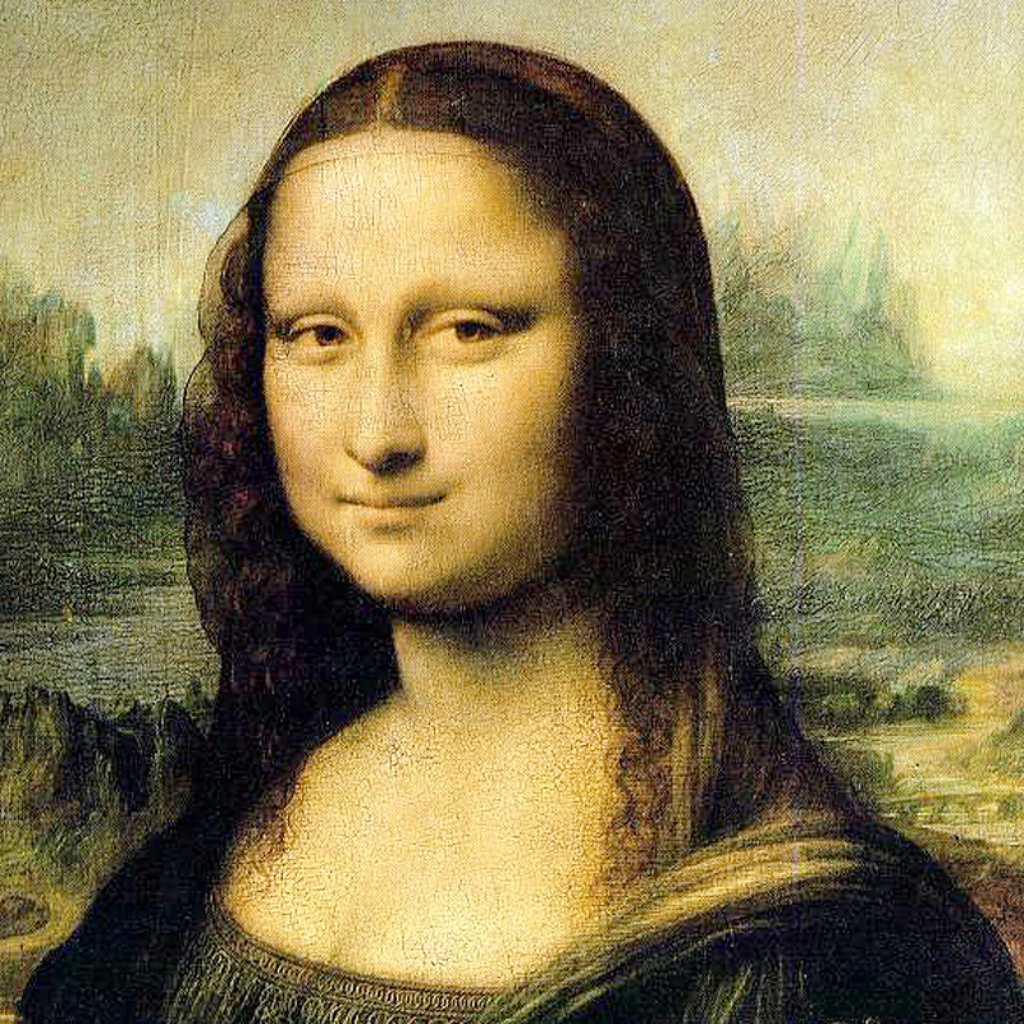}};%
	\end{tikzpicture}%
	\hspace*{-0.3cm}
	\begin{tikzpicture}%
		\node (base) at (0,0) {\includegraphics[width=0.25\linewidth]{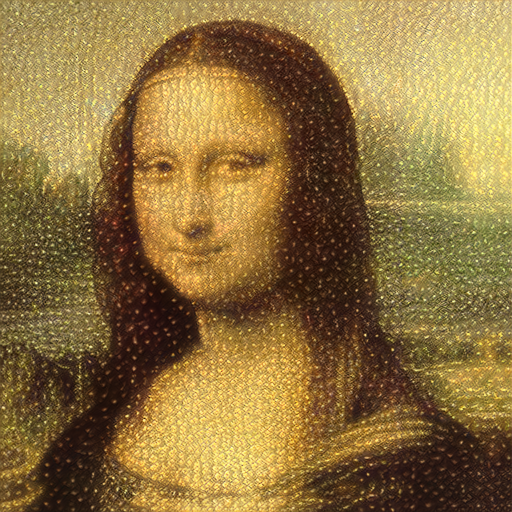}};%
		\node[anchor=north east](overlay) at (base.north east) {\fbox{\includegraphics[width=0.06\linewidth]{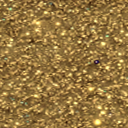}}};%
    \end{tikzpicture}%
	\hspace*{-0.3cm}
	\begin{tikzpicture}%
		\node (base) at (0,0) {\includegraphics[width=0.25\linewidth]{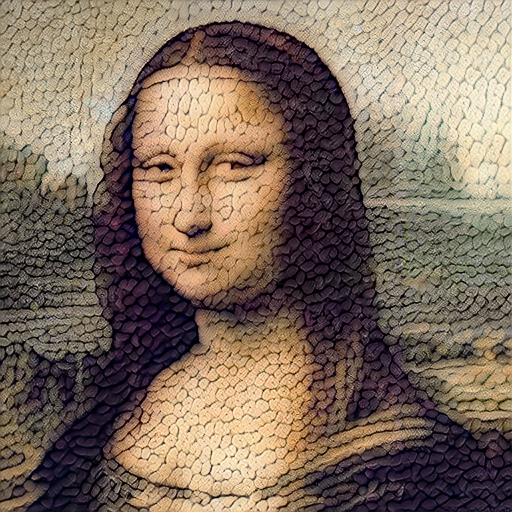}};%
		\node[anchor=north east](overlay) at (base.north east) {\fbox{\includegraphics[width=0.06\linewidth]{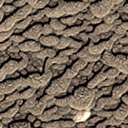}}};%
    \end{tikzpicture}%
	\hspace*{-0.3cm}
	\begin{tikzpicture}%
		\node (base) at (0,0) {\includegraphics[width=0.25\linewidth]{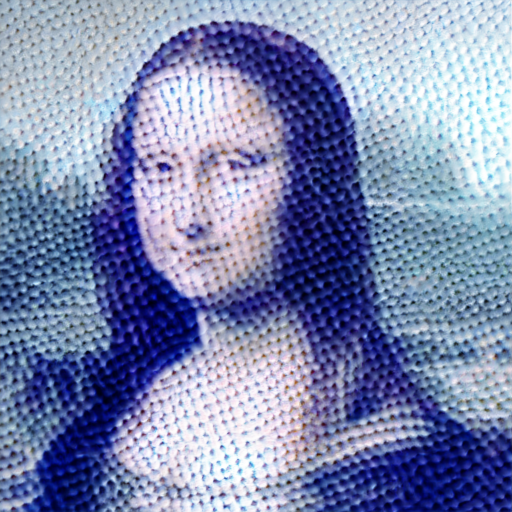}};
		\node[anchor=north east](overlay) at (base.north east) {\fbox{\includegraphics[width=0.06\linewidth]{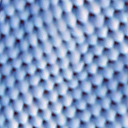}}};%
	\end{tikzpicture}%
	\label{fig:style_transfer:plain}
	}%

	\vspace{-0.4cm}

	\subfloat[Spatial control]{%
	\setlength{\tabcolsep}{1pt}
	\begin{tabular}{cc}
		\begin{tabular}{cc}
			\includegraphics[width=0.12\linewidth]{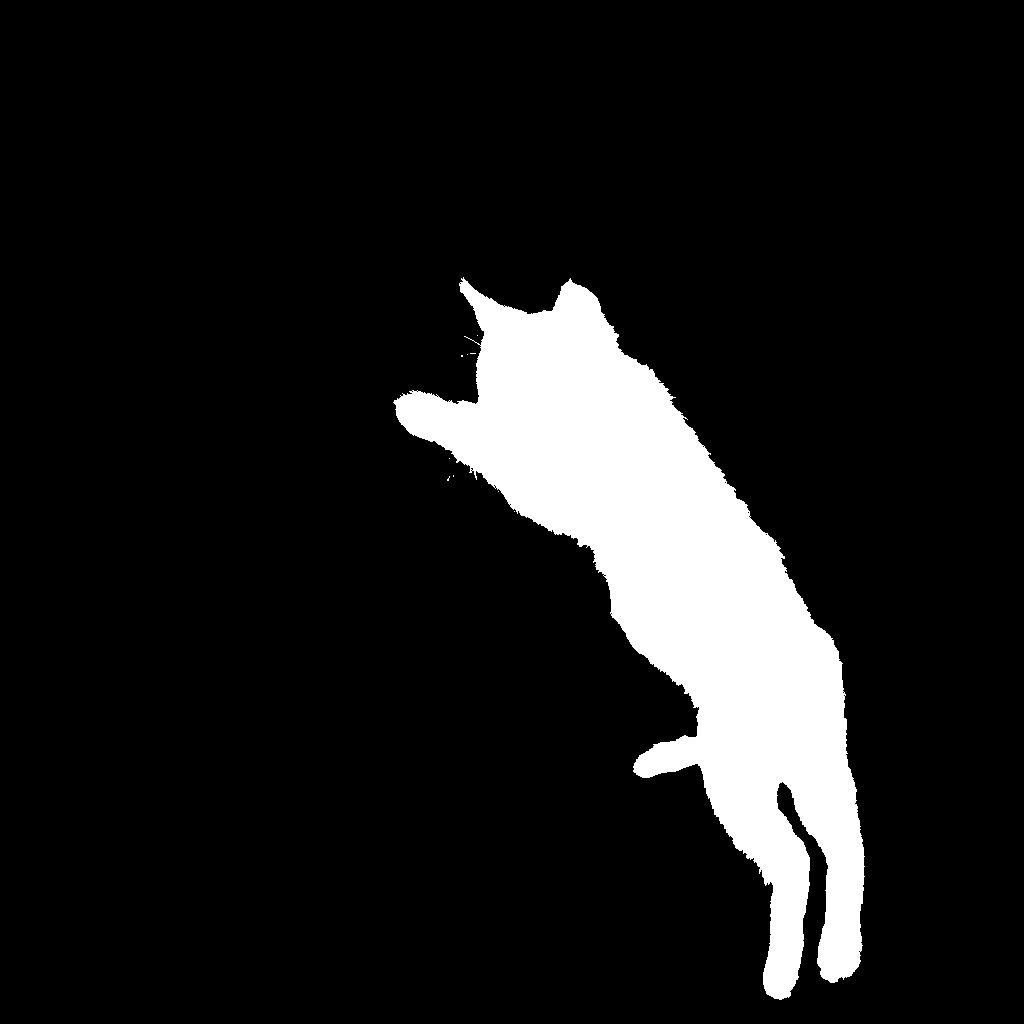} &
			\includegraphics[width=0.12\linewidth]{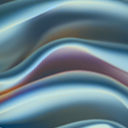} \\[-0.1cm]
			\includegraphics[width=0.12\linewidth]{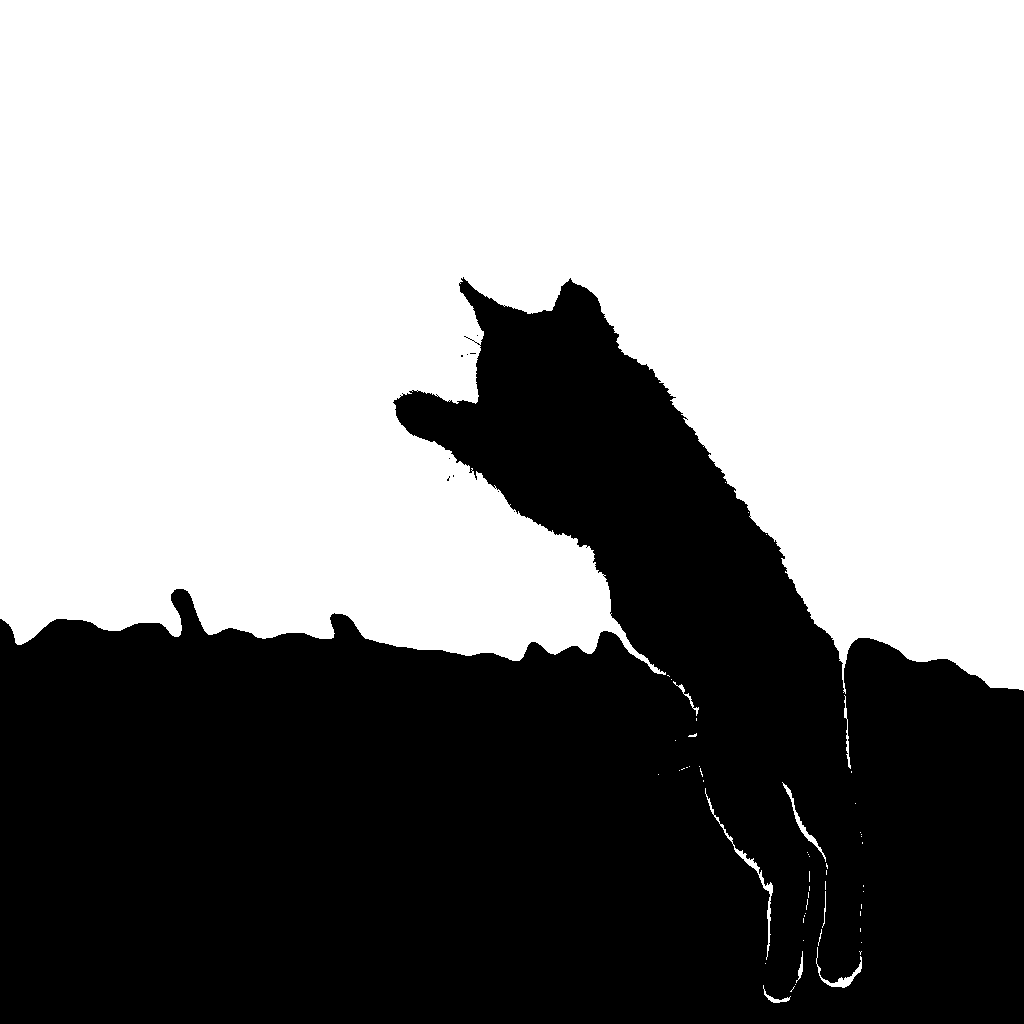} &
			\includegraphics[width=0.12\linewidth]{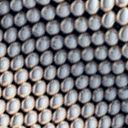}\\[-0.1cm]
			\includegraphics[width=0.12\linewidth]{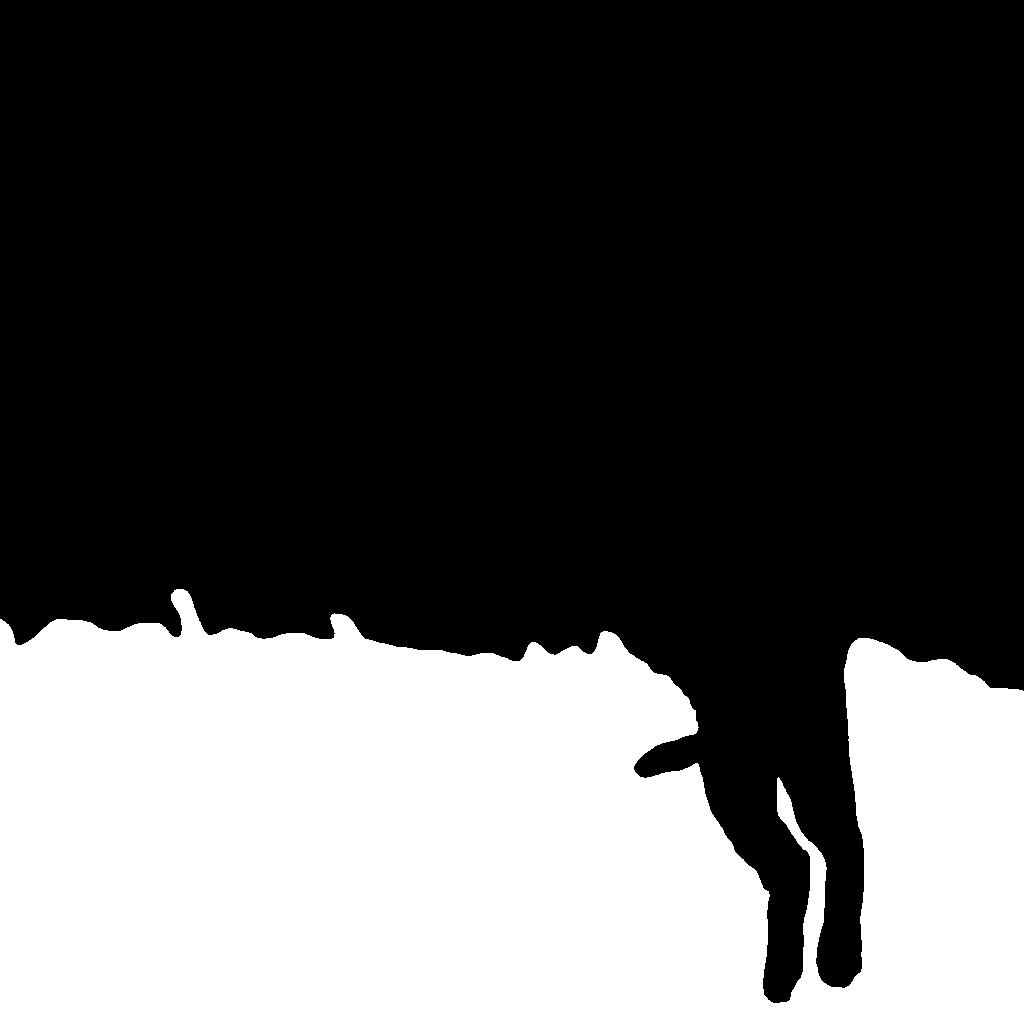} &
			\includegraphics[width=0.12\linewidth]{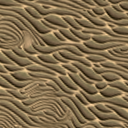}
		\end{tabular} &
		\begin{tabular}{cc}
			\includegraphics[width=0.36\linewidth]{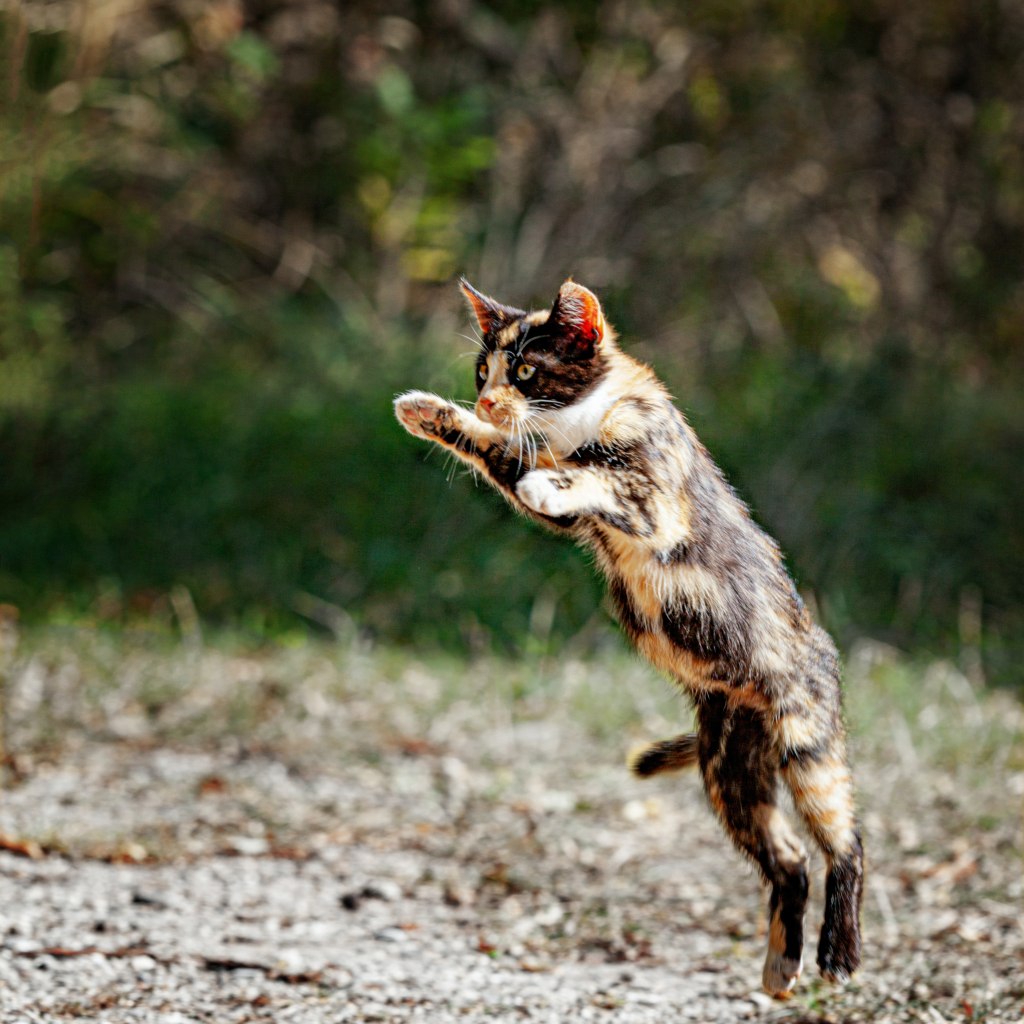} &
			\includegraphics[width=0.36\linewidth]{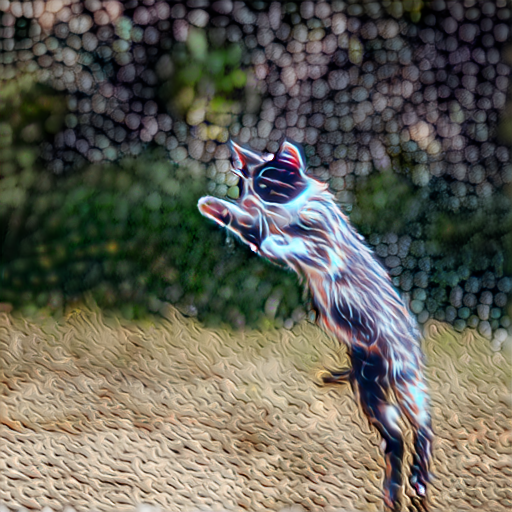}
		\end{tabular}
	\end{tabular}
	\label{fig:style_transfer:spatial}
	}%

	\vspace{-0.4cm}

	\subfloat[Scale control]{%
		\begin{tikzpicture}
			\node (base) at (0,0) {\includegraphics[width=0.25\linewidth]{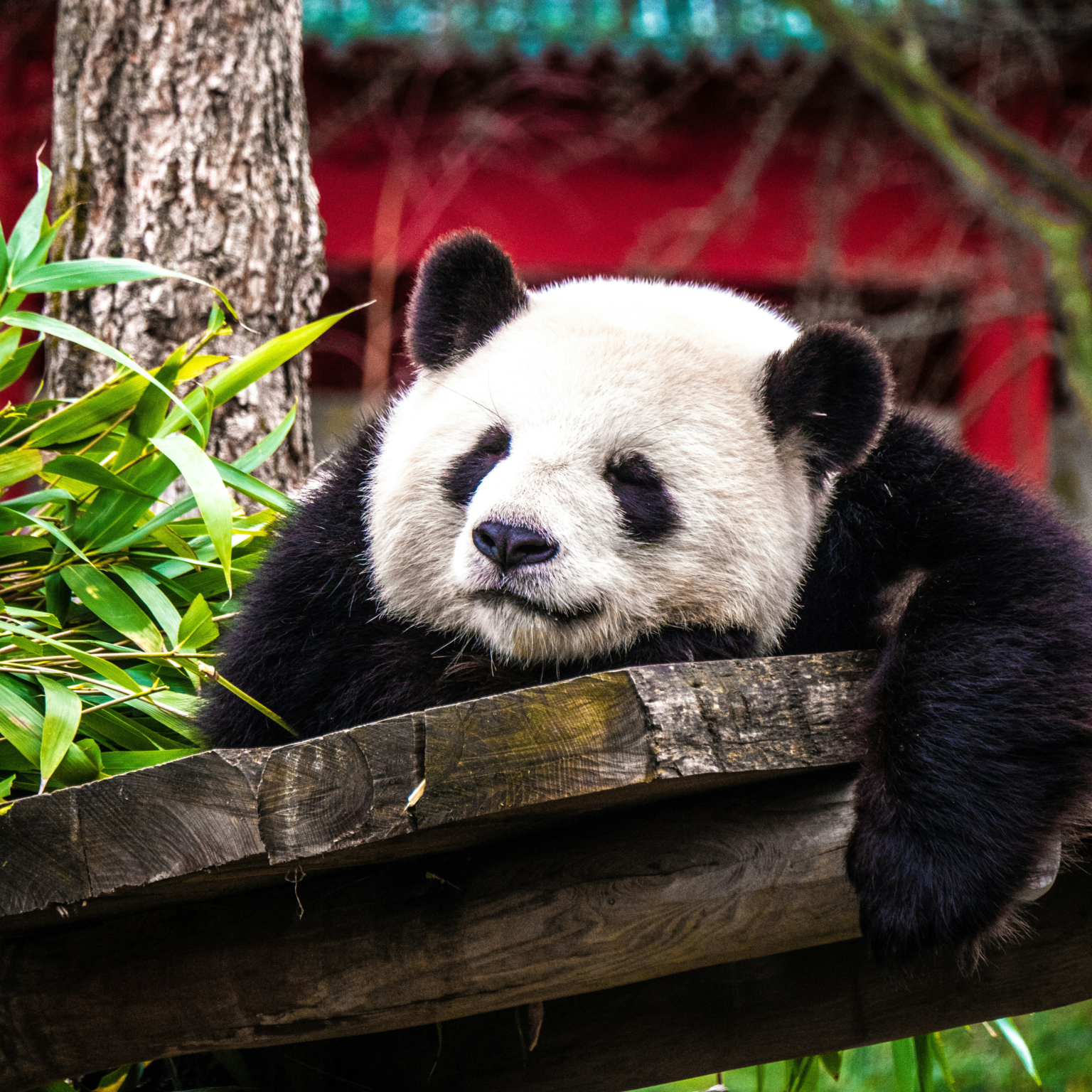}};
		\end{tikzpicture}%
	\hspace*{-0.3cm}
		\begin{tikzpicture}
			\node (base) at (0,0) {\includegraphics[width=0.25\linewidth]{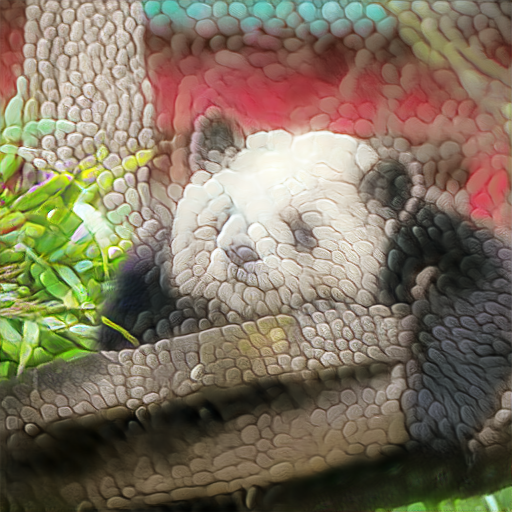}};
			\node[anchor=north east](overlay) at (base.north east) {\fbox{\includegraphics[width=0.06\linewidth]{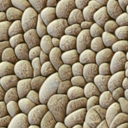}}};
		\end{tikzpicture}%
	\hspace*{-0.3cm}
		\begin{tikzpicture}
			\node (base) at (0,0) {\includegraphics[width=0.25\linewidth]{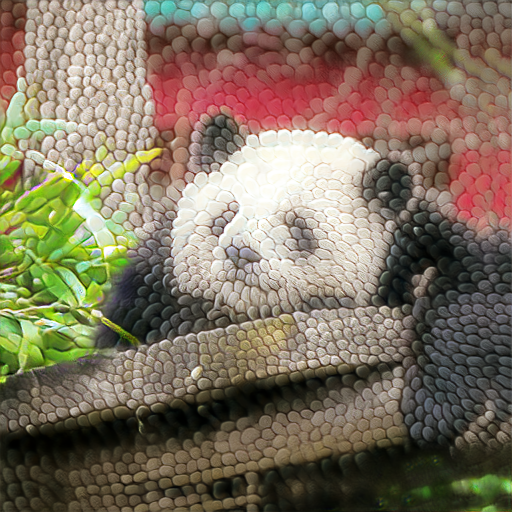}};
		\end{tikzpicture}%
	\hspace*{-0.3cm}
		\begin{tikzpicture}
		\node (base) at (0,0) {\includegraphics[width=0.25\linewidth]{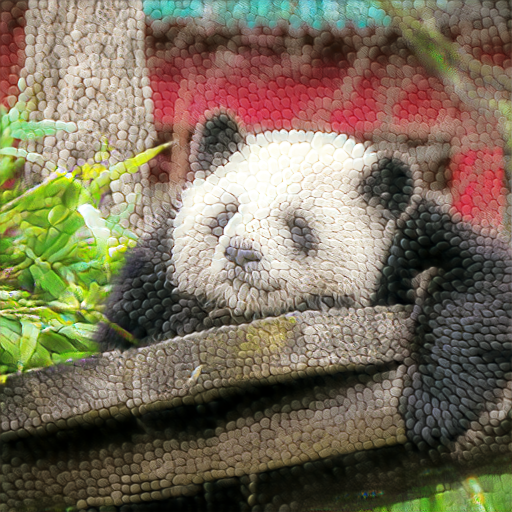}};
		\end{tikzpicture}%
	\label{fig:style_transfer:scale}
	}%

	\caption[]{\textit{Stylizing images by transferring textures.}
		Our approach is adaptable to controllable image stylization. (a) shows plain stylization results without controls.
		(b) and (c) demonstrate spatial and scale controllability of our image stylization method.
		(a) (left) $\copyright$Leonardo da Vinci; (b)(middle) $\copyright$iggii, (c)(left) $\copyright$Chris Curry (\url{unsplash.com}).}
	\label{fig:style_transfer}
\end{figure}

\begin{figure}[hb]
	\centering
	\captionsetup[subfigure]{labelformat=empty}
	
	\subfloat[$\gaussianfeaturevar=0$]{%
	\includegraphics[width=0.19\linewidth]{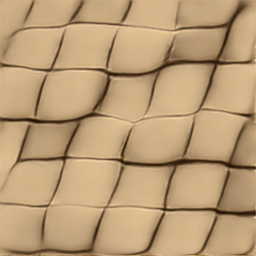}
	}%
	\subfloat[$\gaussianfeaturevar=0.5$]{%
	\includegraphics[width=0.19\linewidth]{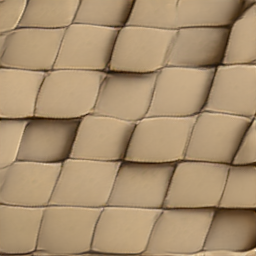}
	}%
	\subfloat[]{%
	\includegraphics[width=0.19\linewidth]{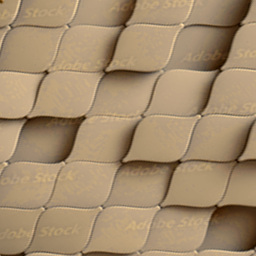}
	}%
	\subfloat[$\gaussianfeaturevar=1.5$]{%
	\includegraphics[width=0.19\linewidth]{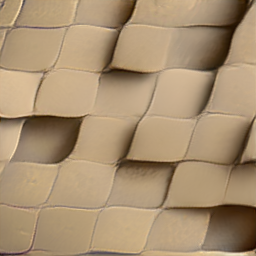}
	}%
	\subfloat[$\gaussianfeaturevar=2$]{%
	\includegraphics[width=0.19\linewidth]{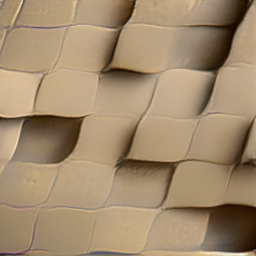}
	}%

    \vspace{-0.3cm}

    \subfloat[$\gaussiancovariancevar=0$]{%
    \includegraphics[width=0.19\linewidth]{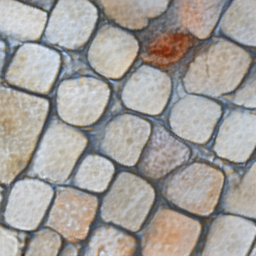}
    }%
    \subfloat[$\gaussiancovariancevar=1/\sqrt{2}$]{%
    \includegraphics[width=0.19\linewidth]{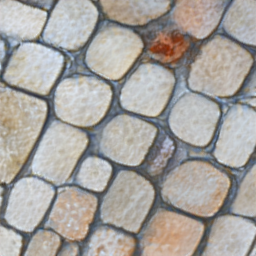}
    }%
    \subfloat[Input]{%
    \includegraphics[width=0.19\linewidth]{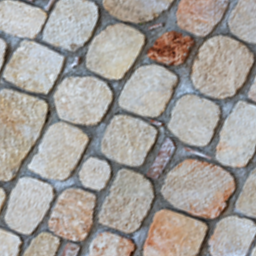}
    }%
    \subfloat[$\gaussiancovariancevar=\sqrt{2}$]{%
    \includegraphics[width=0.19\linewidth]{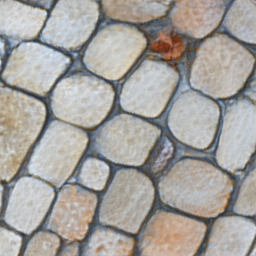}
    }%
    \subfloat[$\gaussiancovariancevar=2$]{%
    \includegraphics[width=0.19\linewidth]{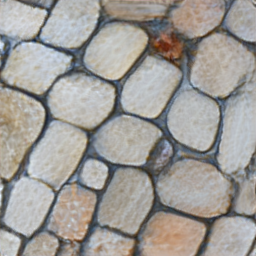}
    }%

	\caption{\textit{Revealing and modifying texture variations.}
	Our method is capable of either reducing or increasing variations in the appearance (top) or geometry (bottom) of a texture by correspondingly changing variations of the appearance or geometry parameters.
	With the input images shown in the center column, we demonstrate reduced variations towards the left and increased variations towards the right.
	$\gaussianfeaturevar$ and $\gaussiancovariancevar$ quantify the variations in appearance and geometry, respectively. See Appendix B.4 for more details. Copyrights: Input (top) $\copyright$RocknRoller Studios; Input (bottom) $\copyright$Vitaliy Sova (Romanization) (stock.adobe.com).}
	\label{fig:variations}
\end{figure}

\begin{figure}
	\centering
	\begingroup
	\setlength{\tabcolsep}{0pt}

	\setcounter{subfigure}{0} %

	\subfloat[source]{%
		\label{fig:texture_interpolation_main:source}%
		\includegraphics[height=0.32\linewidth]{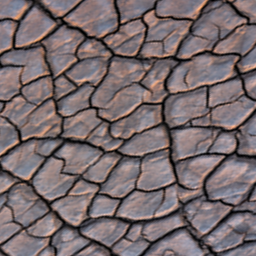}
	}%
	\subfloat[$\interpcoe=0.25$]{%
		\label{fig:texture_interpolation_main:0p25}%
		\includegraphics[height=0.32\linewidth]{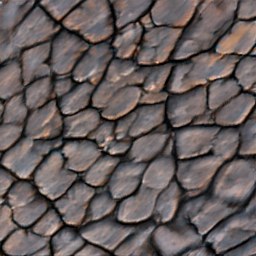}
	}%
	\subfloat[$\interpcoe=0.5$]{%
		\label{fig:texture_interpolation_main:0p5}%
		\includegraphics[height=0.32\linewidth]{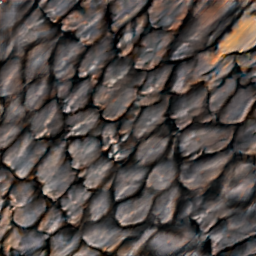}
	}%

	\subfloat[$\interpcoe=0.75$]{%
		\label{fig:texture_interpolation_main:0p75}%
		\includegraphics[height=0.32\linewidth]{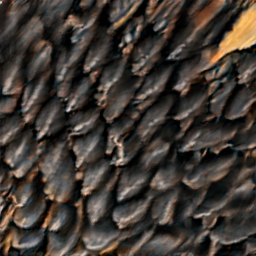}
	}%
	\subfloat[target]{%
		\label{fig:texture_interpolation_main:target}%
		\includegraphics[height=0.32\linewidth]{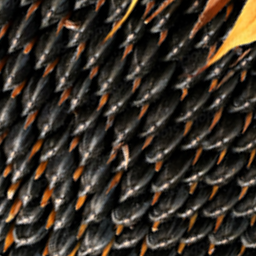}
	}%
	\subfloat[morphing]{%
		\label{fig:texture_interpolation_main:morphing}%
		\includegraphics[height=0.32\linewidth]{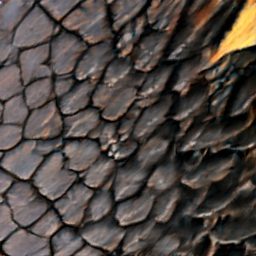}
	}%

	\endgroup
	\caption[]{\textit{Texture interpolation and morphing.} Given a source texture (a) and a target texture (e),
		we can generate interpolations at various degrees between the source and the target,
		controlled by interpolation coefficients of $\interpcoe=0.25$ (b), $\interpcoe=0.5$ (c), and $\interpcoe=0.75$ (d).
		Furthermore, the interpolation coefficients can vary spatially within one output,
		such as linearly increasing from left to right.
		This variation leads to texture morphing, as demonstrated in (f).
		See Appendix B.5 for implementation details.
		(a) $\copyright$Elena Saurius\&Dani Rex/Stocksy; (e) $\copyright$ADDICTIVE STOCK (stock.adobe.com).
	}
	\label{fig:texture_interpolation_main}
	\label{fig:texture_morphing_main}
\end{figure}

\begin{figure}[!htb]
	\centering
	\captionsetup[subfigure]{labelformat=empty}

	\subfloat[Input]{%
	\includegraphics[width=0.25\linewidth]{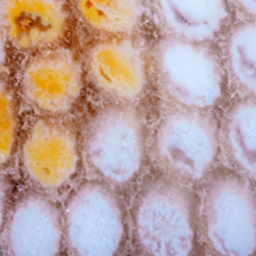}
	}%
	\subfloat[Move]{%
	\includegraphics[width=0.25\linewidth]{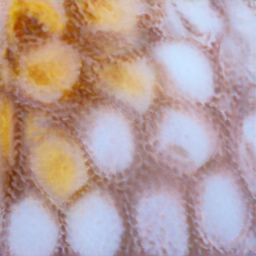}
	}%
	\subfloat[Scale]{%
	\includegraphics[width=0.25\linewidth]{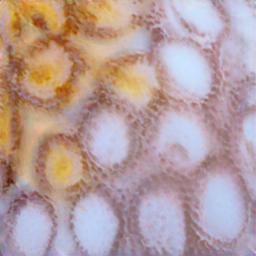}
	}%
	\subfloat[Rotate]{%
	\includegraphics[width=0.25\linewidth]{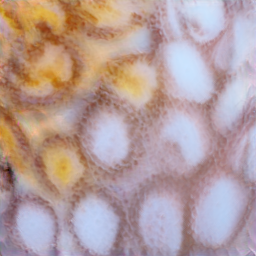}
	}%

	\caption{\textit{Texton editing.} In order to edit individual textons within a texture, we can apply a sequence of edits (move, scale and rotate) to its latent Gaussians. Input: $\copyright$Pansfun Images/Stocksy (stock.adobe.com).
	}
	\label{fig:texton_editing_main}
\end{figure}

\begin{figure}[htb]
	\centering

	\subfloat[]{%
    \includegraphics[width=0.25\linewidth]{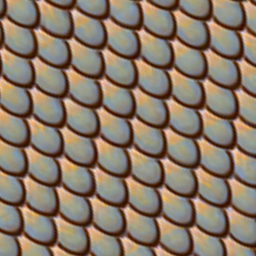}
	}%
	\subfloat[]{%
		\includegraphics[width=0.25\linewidth]{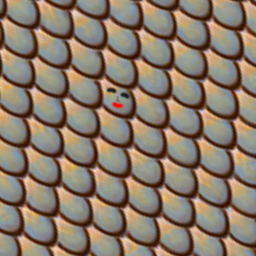}
	}%
	\subfloat[]{%
		\includegraphics[width=0.25\linewidth]{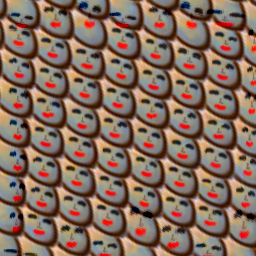}
	}%
	\subfloat[]{%
		\includegraphics[width=0.25\linewidth]{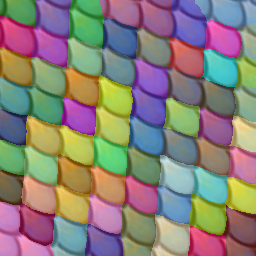}
	}%

	\vspace{-0.8cm}

	\setcounter{subfigure}{0} %

	\subfloat[Input]{%
	\label{fig:edit_prop_main:input}%
	\includegraphics[width=0.25\linewidth]{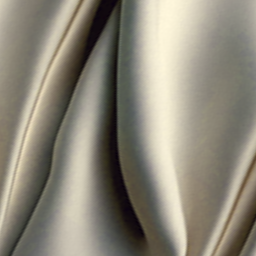}
	}%
	\subfloat[Edited]{%
	\label{fig:edit_prop_main:edited}%
	\includegraphics[width=0.25\linewidth]{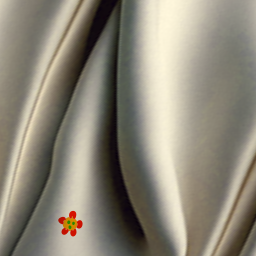}
	}%
	\subfloat[Propagated edits]{%
	\label{fig:edit_prop_main:output}%
	\includegraphics[width=0.25\linewidth]{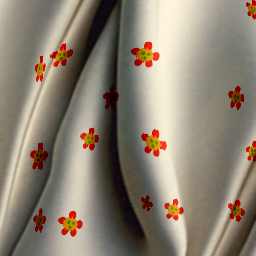}
	}%
	\subfloat[Seg. overlay]{%
	\label{fig:edit_prop_main:seg}%
	\includegraphics[width=0.25\linewidth]{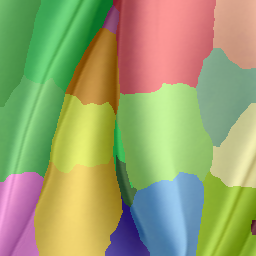}
	}%

	\caption[]{\textit{Edit propagation.}  Given a texture (a), we can edit it (b) and the edits can be propagated to other similar locations (c) based on extracted textons (d), for both stationary (top) and non-stationary (bottom) structures.}
	\label{fig:edit_prop_main}
\end{figure}

\section{Applications}
\label{sec:applications}

We showcase various applications for our compositional neural texture representation.
\note{
including texture diversification (Section~\ref{sec:diversification}),
texture transfer and image stylization (Section~\ref{sec:transfer}),
modifying texture variations (Section~\ref{sec:application:variations}),
texture interpolation (Section~\ref{sec:interpolation}),
direct manipulation of textons (Section~\ref{sec:interface}),
and \replace{the visualization of flow fields}{animated textures} (Section~\ref{sec:animation}).
}%
Each application only involves encoding input textures as latent Gaussians $\setof{\gaussian}{\gaussianindex}_{\gaussianindex=1}^{\numgaussians}$ and editing $\setof{\gaussian}{\gaussianindex}_{\gaussianindex=1}^{\numgaussians}$, and the output textures are generated efficiently by feeding the edited latent Gaussians through the generator network in a feed-forward manner.
{\em We don't require any additional training or fine-tuning for each application}.
For implementation details and more results of each application, please refer to Appendix B.
For efficiency reference, the feed-forward processing of a $256\times256$ image using our autoencoder takes 0.16 seonds on a single NVIDIA A10G (24GB) in PyTorch,
and this is achieved without any specific optimizations for speed.

\paragraph{Texture Diversification}
\label{sec:diversification}

By {\it appearance feature reshuffling} among valid textons (with rounded $\gaussianweightsymbol=1$ during inference), we can create diverse versions of the same texture, as shown in \replace{Figure~\ref{fig:diverse_texture_synthesis:output1,fig:diverse_texture_synthesis:output2,fig:diverse_texture_synthesis:output3}}{Figure~\ref{fig:diverse_texture_synthesis_main}}.
\note{
This method validates that our neural textons accurately encapsulate the texture's composition, showing that feature reshuffling creates a subspace of variations of a texture.
}%

\paragraph{Texture Transfer}
\label{sec:transfer}

\replace{Texture transfer combines the structure and appearance of two different textures: one provides the structure while the other provides the appearance.
}{%
Texture transfer combines the structure of one texture with the appearance of another (Figure~\ref{fig:qualitative_texture_transfer}).
}%
\note{Our compositional texture representations effectively separate the spatial structure and appearance characteristics within a texture, 
which is key for successful texture transfer.
}%
Our results are created by taking the Gaussians from the structure-providing (SP) texture, appearance features from the appearance-providing (AP) texture,
and we \new{only} adjust the mean of Gaussian appearance features of the SP texture to match that of the AP texture.
\note{
We can also directly replace appearance features from structure-providing textures with those from appearance-providing textures. Please see Section~\ref{sec:appendix:transfer} for more results and discussions.
}%

\paragraph{Image stylization}
We can stylize a natural image using textures.
While our network is not trained on natural images, a small patch of a natural image can be effectively treated as a texture.
Figure~\ref{fig:style_transfer:plain} shows the results by transferring textures to a content image patch by patch.
Users can also transfer different textures to image regions specified by control masks, as shown in Figure~\ref{fig:style_transfer:spatial}.
\note{, which could be later used to create interactive brush tools.
}%
Moreover, Figure~\ref{fig:style_transfer:scale} shows scale control over transferred textons, achieved by varying the scale of Gaussians derived from content image patches.

\note{
Figure~\ref{fig:style_transfer} shows the transfer of three different textures to three image regions.
Moreover, users can control the scale of transferred textons, as demonstrated in Figure~\ref{fig:style_transfer} and more in Section~\ref{sec:appendix:transfer}.
}%

\note{manually create the structure the textons, and then transfer the appearance. draw SIGGRAPH}

\paragraph{Revealing and Modifying Texture Variations}
\label{sec:application:variations}

We can modify non-local texture variations by editing variations within the Gaussian's appearance and geometry parameters.
To edit the variations, we first calculate the means of Gaussian appearance and geometry parameters and edit their deviations from the means.
Figure~\ref{fig:variations} show the edited image when we reduce and increase the variations in appearance and geometry parameters respectively.

\paragraph{Texture Interpolation}
\label{sec:interpolation}

Our representation can be applied for interpolating textures by interpolating latent Gaussians, such as their spatial positions, features and weights.
The interpolated texture is reconstructed from the interpolated Gaussians\note{$\{\gaussianinterp\}$}.
\replace{Figure~\ref{fig:texture_interpolation}}{Figures~\ref{fig:texture_interpolation_main:source} to \ref{fig:texture_interpolation_main:target}} show interpolated textures with various interpolation coefficients $\interpcoe$.
Our method can also generate smooth transition of two textures in one image (Figure~\ref{fig:texture_interpolation_main:morphing}) with spatially varying interpolation coefficient.

\paragraph{Direct texton manipulation}
\label{sec:interface}

\note{
Since we extract individual neural textons, our approach can also enable more direct manipulation of individual textons.

\subsubsection{Editing Gaussians}
\label{sec:interface:gaussian_editor}
}%

Figure~\ref{fig:texton_editing_main} shows users can move, scale and rotate textons by editing their geometric parameters ($\gaussianmean$,$\gaussiancovariance$,$\gaussiandir$).

\paragraph{Edit propagation}

When users make edits to a specific area of a texture,
these changes can be automatically propagated to other similar areas. This process is demonstrated in Figure~\ref{fig:edit_prop_main}.

\paragraph{\replace{Flow visualization}{Texture Animation}}
\label{sec:animation}

\note{
In Section~\ref{sec:interface:gaussian_editor},
we describe how our framework can handle image editing at the texton level, with each latent Gaussian aligned with a corresponding texton in the image.
}%
Our method enables animating Gaussians using flow fields \cite{Kwatra:2005:TOE}\note{\cite{Chen:2011:DTV}}, which are then reconstructed as animated textures. Please see our supplementary video.

\begin{figure}[htb]
	\centering
	\captionsetup[subfigure]{labelformat=empty}

	\note{
	\subfloat[]{%
	\includegraphics[width=0.3\linewidth]{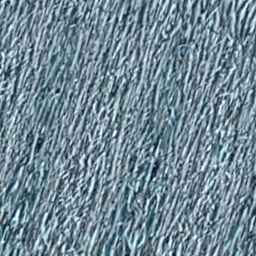}
	}%
	\subfloat[]{%
	\includegraphics[width=0.3\linewidth]{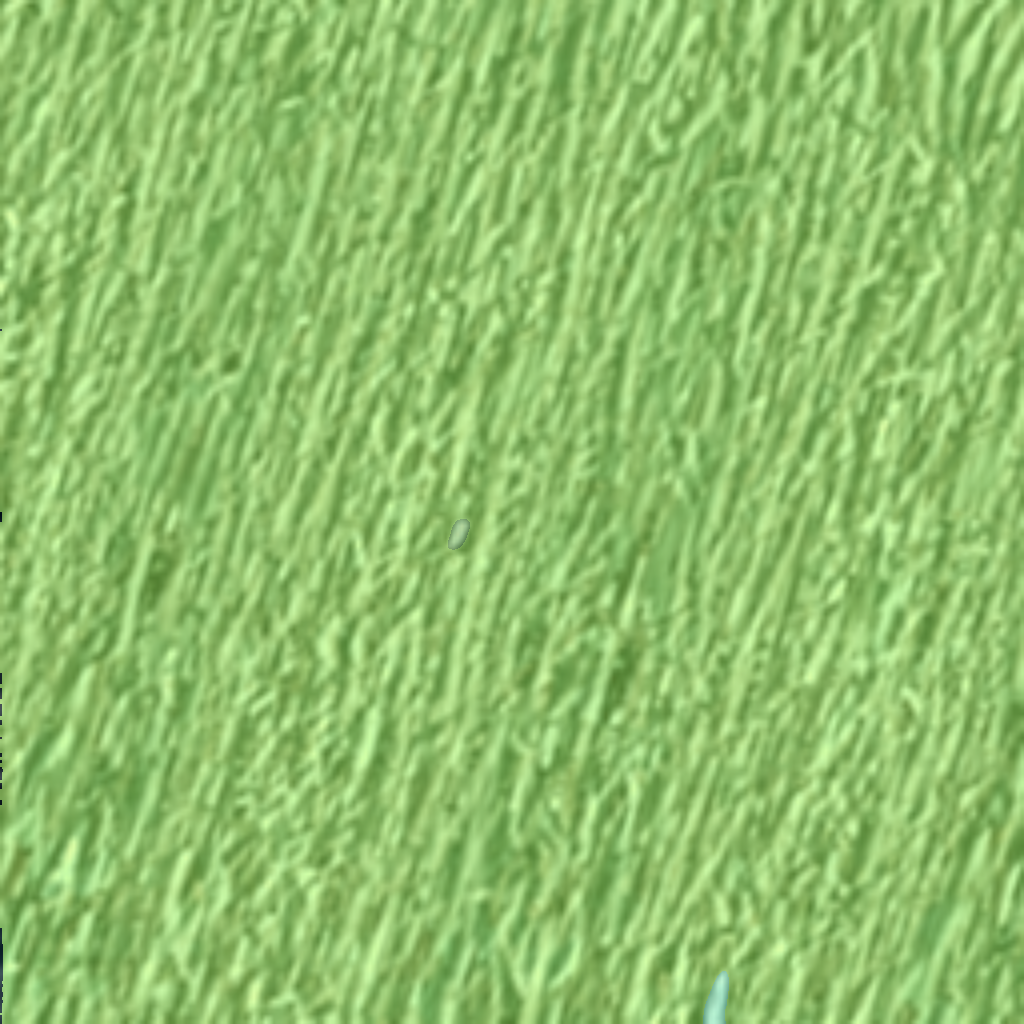}
	}%
	\subfloat[]{%
	\includegraphics[width=0.3\linewidth]{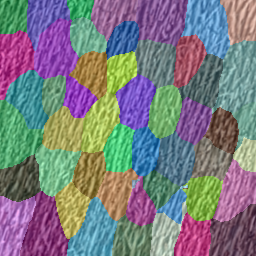}
	}%

	\vspace{-0.8cm}%
	}%

	\subfloat[]{%
	\includegraphics[width=0.3\linewidth]{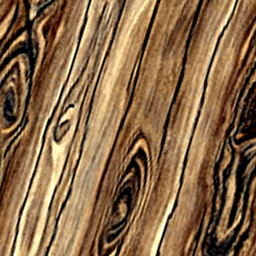}
	}%
	\subfloat[]{%
	\includegraphics[width=0.3\linewidth]{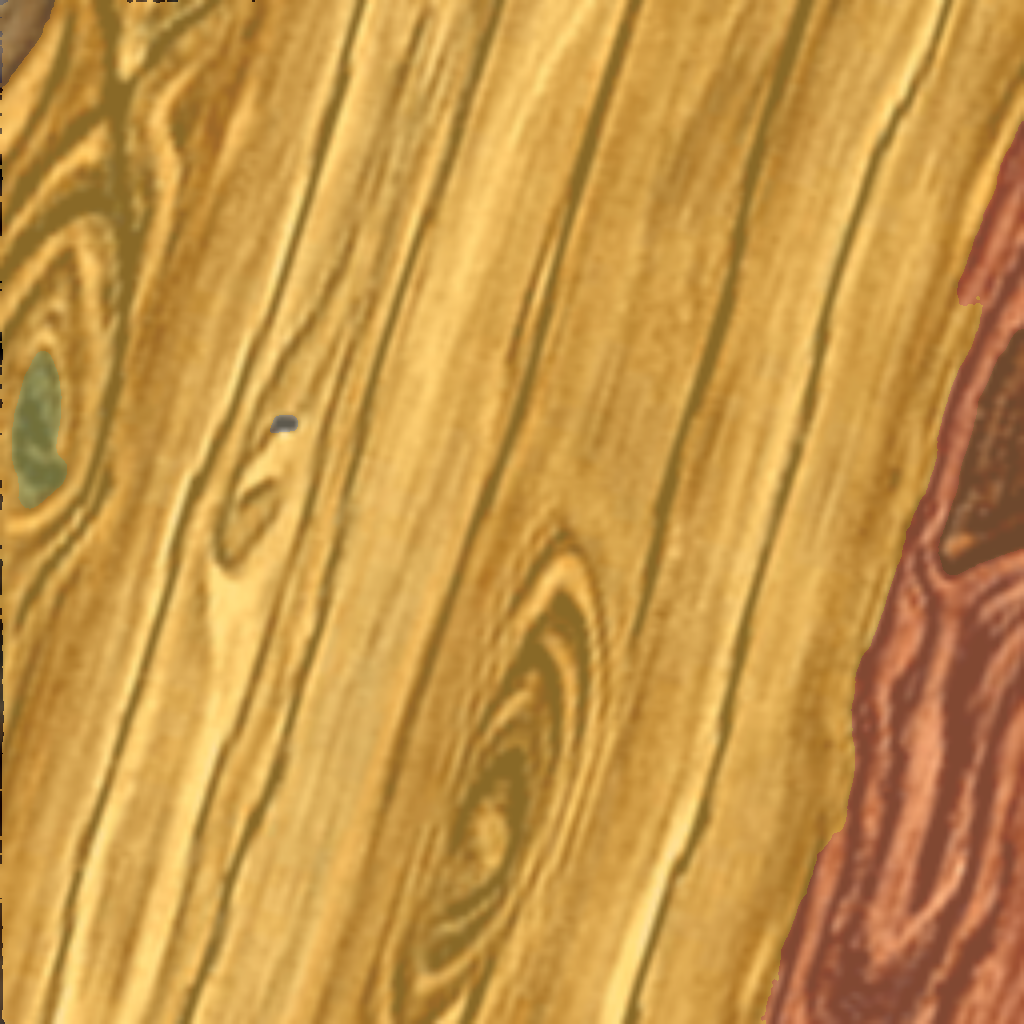}
	}%
	\subfloat[]{%
	\includegraphics[width=0.3\linewidth]{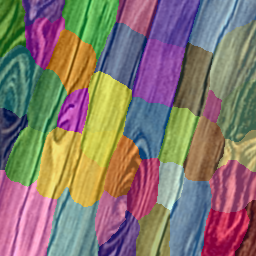}
	}%

	\vspace{-0.8cm}%

	\subfloat[Input]{%
	\includegraphics[width=0.3\linewidth]{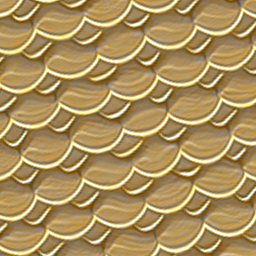}
	}%
	\subfloat[\cite{Meta:2023:SA}]{%
	\includegraphics[width=0.3\linewidth]{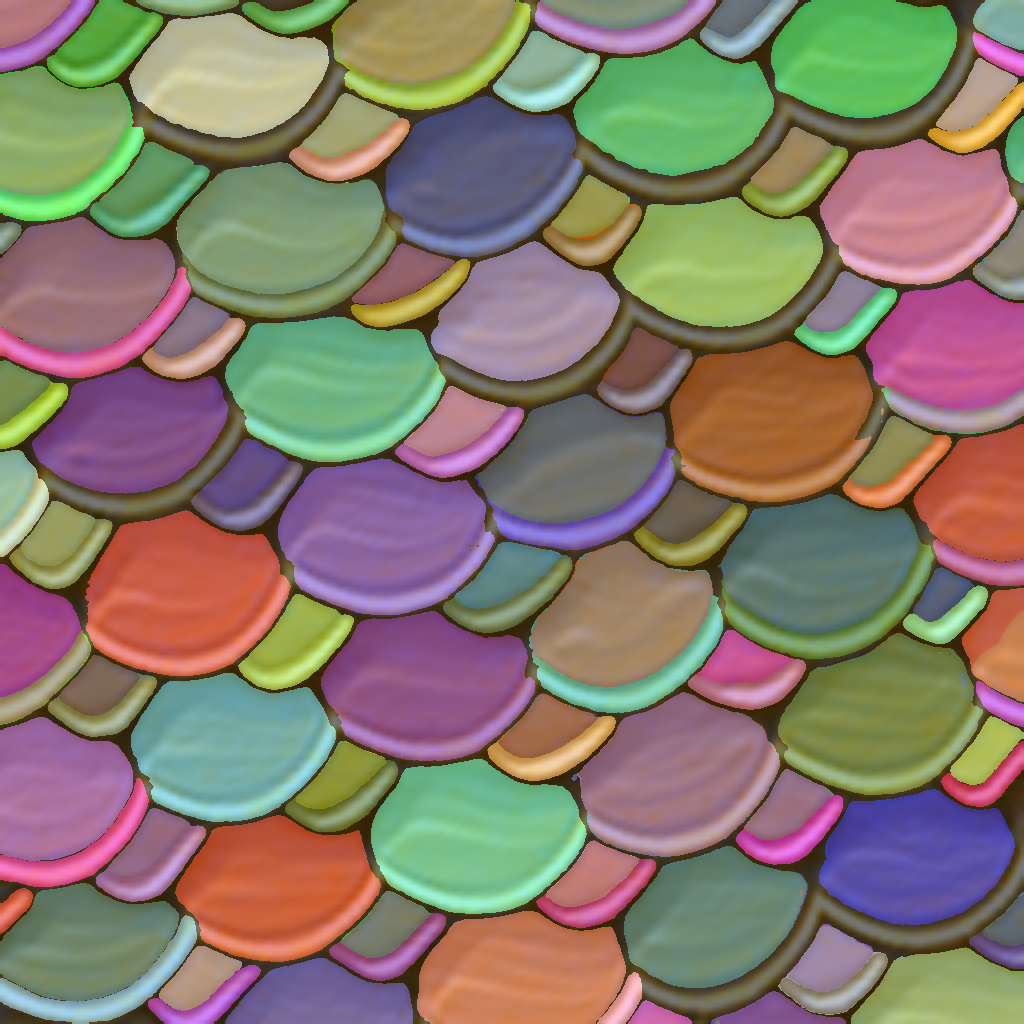}
	}%
	\subfloat[Ours]{%
	\includegraphics[width=0.3\linewidth]{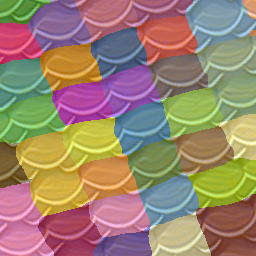}
	}%

	\caption[]{\textit{Comparison to segment anything (SA).}
		SA is trained with supervision of 1 billion masks. However, SA is incapable of detecting and extracting textons from textures.
		Our model can extract textons from textures using fully unsupervised training.
	}
	\label{fig:comparison_segment}
\end{figure}

\begin{figure}[h]
	\centering
	\captionsetup[subfigure]{labelformat=empty}
	\subfloat[Structure]{\includegraphics[width=0.32\linewidth]{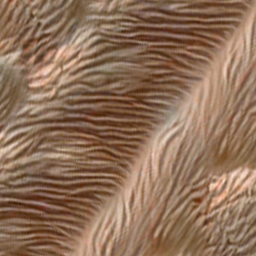}}
	\subfloat[Appearance]{\includegraphics[width=0.32\linewidth]{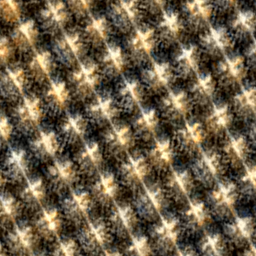}}
	\subfloat[\cite{Kolkin:2019:STR}]{\includegraphics[width=0.32\linewidth]{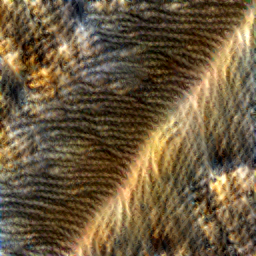}}
	
	\subfloat[\cite{Tumanyan:2022:SVF}]{\includegraphics[width=0.32\linewidth]{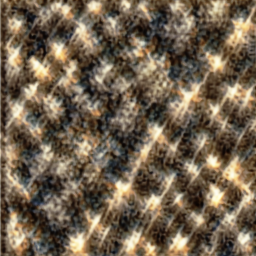}}
	\subfloat[\cite{Cheng:2023:GII}]{\includegraphics[width=0.32\linewidth]{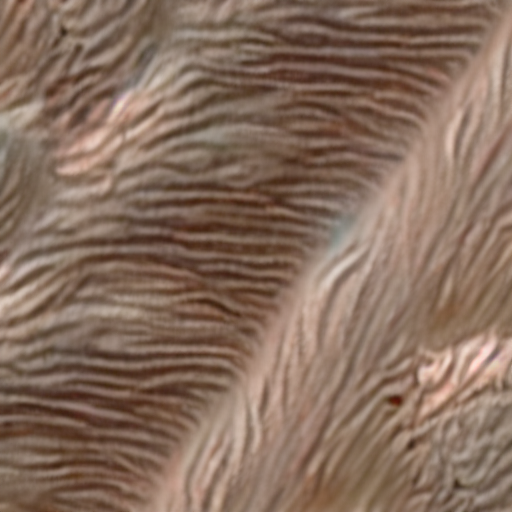}}
	\subfloat[ours]{\includegraphics[width=0.32\linewidth]{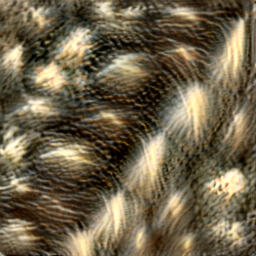}}

	\caption[]{\textit{Comparisons with existing approaches on texture transfer.} Our approach can preserve the better structure from the structure-providing texture than \cite{Tumanyan:2022:SVF} and more accurate texton appearance than \cite{Kolkin:2019:STR}. 
		Cheng et al.~\shortcite{Cheng:2023:GII} could not transfer the texton appearance to the structure image due to its use of text-to-image models while it might not be suitable to describe detailed textures using abstract human language.
	}
	\label{fig:comparison_texture_transfer_main}
\end{figure}

\section{Analysis}
\label{sec:analysis}

We compare our method to related approaches and evaluate our approach via ablation studies, both qualitatively and quantitatively.

\paragraph{Comparison to ``Segment Anything''}
\label{sec:comparison_segment_anything}
To evaluate the quality of our extracted textons, we compare our segmentation produced by $\segmentationnetwork$ against the ``Segment Anything'' (SA) model \cite{Meta:2023:SA}.
The SA model has been trained on a dataset of 1 billion masks with remarkable ability to segment images with zero prior knowledge.
Figure~\ref{fig:comparison_segment} shows the comparisons with the SA model. See Appendix C.1 for more comparisons.
These results indicate SA is incapable of reliably extracting textons, a fundamental aspect that contributes to the versatility of our method across a variety of texture applications.

\paragraph{Comparisons on Texture Transfer}

\note{In this section, We aim to demonstrate our approach's unique capacity in disentangled structure-appearance texture modeling using latent Gaussians based on segmented structures.}%
\note{To this end, }We compare our approach with prior methods \cite{Kolkin:2019:STR,Tumanyan:2022:SVF,Cheng:2023:GII} tailored for combining appearance from one image and content/structure from another image.
\replace{Figure~\ref{fig:comparison_texture_transfer}}{Figure~\ref{fig:comparison_texture_transfer_main}} shows the comparisons.
Our method generates more visually appealing results without clear artifacts,
effectively capturing both the structure and appearance from respective SP and AP textures, demonstrating the unique capacity of our approach in disentangled structure-appearance texture modeling.
Note we not only disentangle structure from texture appearance but also explicitly represent them as different Gaussian parameters, enabling various applications (Section~\ref{sec:applications}) beyond texture transfer.
See Appendix C.2 for more comparisons.

\paragraph{Ablation Studies}
\label{sec:analysis:ablation}

We qualitatively and quantitatively evaluate our method via ablation studies to demonstrate the importance of the method components, including loss functions and architectures.
The ablation studies suggest our combined design of loss functions and architectures are crucial for the overall performance of our method. See Appendix C.3 to C.5 for details about the ablation studies.

\note{
}%

\section{Limitations and Future Work}
\label{sec:limitation}
\label{sec:conclusion}

\begin{figure}
\centering
\captionsetup[subfigure]{labelformat=empty}
\captionsetup[subfigure]{justification=centering}

\subfloat[Input]{%
    \label{fig:limitation:elongated:input}
    \includegraphics[width=0.24\linewidth]{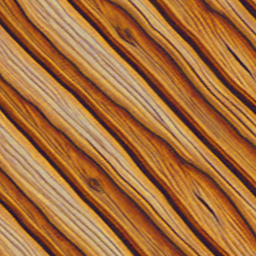}
}%
\subfloat[Gaussians]{%
    \label{fig:limitation:elongated:gaussian}
    \includegraphics[width=0.24\linewidth]{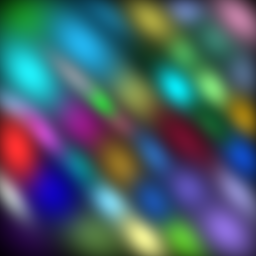}
}%
\subfloat[Edited (rotated) Gaussians]{%
    \label{fig:limitation:elongated:edited_gaussian}
    \includegraphics[width=0.24\linewidth]{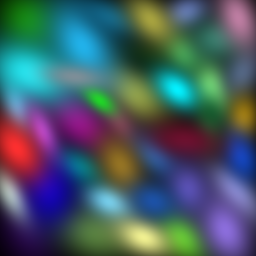}
}%
\subfloat[Edited]{%
    \label{fig:limitation:elongated:edited}
    \includegraphics[width=0.24\linewidth]{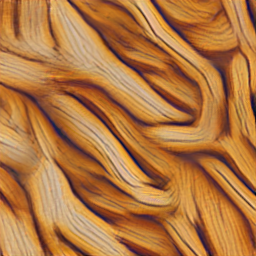}
}%
\caption[]{\textit{Improper compositional representation for elongated structures.}
    For elongated, continuous structures, it might be improper to model them as localized textons.
    Editing individual textons might discrupt the continuous structures.
}
\label{fig:limitation:elongated}
\end{figure}

Our approach relies on the assumption of random reshuffling of texton appearances ({\scshape Property~\ref{property:reshuffle}} ({\scshape Reshuffling Invarance})) that can result in a different version of the original texture without altering its overall appearance.
Consequently, while our method can extract non-stationary structure and generate spatially-varying textures during inference (e.g. Figures~\ref{fig:style_transfer:spatial} and \ref{fig:texture_interpolation_main:morphing}), \replace{it is suitable for those}{the network is \replace{trained on}{inherently more suitable for} textures} that exhibit consistent texton appearances throughout the image domain.

Our current method may struggle with patterns with elongated, continuous structures.
In these patterns, the notion of textons might not be clear and each elongated structure might be represented by multiple textons in our current implementation as we impose {\scshape Property~\ref{property:compact}} ({\scshape Spatial Compactness}) on \replace{these learned neural}{learned} textons.
Figure~\ref{fig:limitation:elongated} shows such an example where texton editing can disrupt continuous structures.
\finalnew{Our approach might not capture the largest possible textons, \replace{where a single texton represents an entire elongated structure. This larger representation}{which} would be more aligned with user intentions during editing.
}%
\finalnew{However, in applications such as texture transfer, this limitation is less observable, as shown in Figure 17 (3rd and 4th columns) in the appendix, because the sequences of Gaussians that encode elongated or large structures remain spatially intact during appearance transfer.
}%
Although textons are ill-defined with ambiguity \finalnew{especially for highly unstructured or elongated textures (e.g. textons could be single hair or strands of hair in hair textures)}, the practical utility of the extracted textons on various textures using the 4 properties (Section~\ref{sec:method:properties}) is evident by the various applications.

\finalnew{The Gaussian representation may lose fine details, such as sharp edges or corners due to the smoothness nature of Gaussian splatted feature maps and the limited number of Gaussians used. The texture representation has a focus on editing capability with a trade-off versus modeling capability.
Extending the single-level representation to a multi-level representation (e.g., Gaussians of different sizes for different levels) could be an interesting future research direction to better preserve editing capability while increasing modeling capability, especially for sharp texture features.
}%
\finalnew{In addition, as we model a texture as a set of textons that can fully cover the texture, spatial manipulation may expose undefined background to cause extra artifacts.
Therefore, another interesting future direction is to model a texture as textons in the foreground plus a background to improve both editing and modeling capabilities.
}%
Finally, our approach could incorporate supervised training signals from pre-existing semantic segmentation models (e.g. segment anything) to extract semantic textons more accurately.

In this paper, we propose a framework for compositional modeling of repetitive \replace{patterns in images}{textures}.
Using our framework, we present various texture editing applications using simple, baseline approaches, which could be improved with more dedicated mechanisms in the future.
For example, in texture interpolation applications (Figure~\ref{fig:texture_interpolation_main}),
we could interpolate Gaussians via a softer mechanism using optimal transport instead of via the hard approach by solving assignment problems (see Appendix B.5).
Additionally, our approach could extend beyond the presented use cases,
such as texture rectification \cite{Hao:2023:DHT},
by analyzing and rectifying Gaussian centers, directions and shapes,
and weathering \cite{Bellini:2016:TVW},
by analyzing and manipulating the appearance features,
font stylization by focusing on font shape, and video stylization by imposing intuitive coherence directly on latent Gaussians.
Furthermore, our framework could be adapted to model repetitive phenomena in other data types, such as materials, vector patterns, 3D geometry, and audio.

\begin{acks}
We would like to thank the anonymous reviewers for their valuable feedback. This work was supported in part by NSF grant No. IIS2126407.
\end{acks}

\bibliographystyle{ACM-Reference-Format}
{

}

\end{document}